\documentclass[pre,aps,twocolumn,superscriptaddress]{revtex4}

\usepackage[dvips]{graphicx}
\usepackage{amssymb,amsfonts,amsmath}
\usepackage{color}
\usepackage{ulem}
\usepackage{slashed}

\newcommand{\rhotwo}{\rho^{(2)}}
\newcommand{\rv}{{\bf r}}

\begin{document}

\title{Mean-Field Theory of Inhomogeneous Fluids}

\author{S.M. Tschopp}
\affiliation{Department of Physics, 
  University of Fribourg, CH-1700 Fribourg, Switzerland}

\author{H.D. Vuijk}
\affiliation{Leibniz-Institut f\"ur Polymerforschung Dresden, 
Institut Theorie der Polymere, 01069 Dresden, Deutschland}

\author{A. Sharma}
\affiliation{Leibniz-Institut f\"ur Polymerforschung Dresden, 
Institut Theorie der Polymere, 01069 Dresden, Deutschland}

\author{J.M. Brader}
\address{Department of Physics, 
  University of Fribourg, CH-1700 Fribourg, Switzerland}

\begin{abstract}
The Barker-Henderson perturbation theory is a bedrock of liquid-state physics, 
providing quantitative predictions for the bulk thermodynamic properties 
of realistic model systems. 
However, this successful method has not been exploited for the study of 
inhomogeneous systems. 
We develop and implement a first-principles `Barker-Henderson density functional', thus providing 
a robust and quantitatively accurate theory for classical fluids in external fields.  
Numerical results are presented for the hard-core Yukawa model in three dimensions.
Our predictions for the density around a fixed test particle and between planar walls 
are in very good agreement with simulation data. 
The density profiles for the free liquid vapour interface show the expected 
oscillatory decay into the bulk liquid as the temperature is reduced towards the 
triple point, but with an amplitude much smaller than that predicted 
by the standard mean-field density functional. 
\end{abstract}


\maketitle


\section{Introduction}
In 1873 van der Waals presented his celebrated equation of state, 
which corrected the well-known ideal gas expression to account for the influence 
of interparticle interactions \cite{vdw}.  
The key physical insight, nowadays fundamental 
to mean-field and perturbation theories of classical fluids, is the separation 
of the two effects of particles occupying a certain volume, due to their 
mutual repulsion, and of them attracting 
each other. 
The validity of this separation rests on the assumption that the attractive 
component of the pair interaction is both weak and long-ranged, as was 
pointed out by Boltzmann in 1895 \cite{boltzmann}, Ornstein in 1908 \cite{ornstein}, 
and only much later proved rigorously by Kac and coworkers in the 1960's 
\cite{kac}. 
If these conditions are satisfied then one arrives at a physically intuitive picture 
in which the average microstructural arrangement of the particles in a liquid, as 
characterized by spatial correlation functions, is largely determined by strongly repulsive 
short-range interaction forces, with the long-range attractive forces exerting 
only a perturbing influence. 
The system interacting via the purely repulsive part of the pair potential provides 
a reference or starting point for the description of realistic liquid models, 
thus playing a role analogous to that of the harmonic lattice for the development of 
theories of solids.

The first step in turning the approach of van der Waals into a modern statistical mechanical 
theory of  liquids was taken by Zwanzig \cite{zwanzig}. 
In 1954 he showed how an attractive component to 
the pair interaction potential (he considered a square-well attraction) 
could be treated systematically  
using perturbation theory, an approach sometimes referred to as the `high-temperature 
expansion', because the expansion parameter is the 
attractive part of the potential scaled by $k_BT$. 
However, at that time there was no adequate theory of the repulsive reference system, so the 
method found little immediate application.
A key step was the development of an acceptably accurate and, importantly, analytically 
tractable theory of the hard-sphere system, the Percus-Yevick theory of 1958 
\cite{py}. 
The remarkable analytical solution of this approximate closure to the Ornstein-Zernike 
integral equation provided closed form expressions for both thermodynamic quantities and 
pair correlation functions \cite{wertheim,thiele,smith}. 

In a seminal pair of papers from 1967 Barker and Henderson combined the approach 
of Zwanzig with the Percus-Yevick results for hard-spheres to obtain the first true 
microscopic theory of liquids, embedding the ideas of van der Waals within the framework 
of statistical mechanics \cite{bh_original1,bh_original2} (reviewed in \cite{bh_review}).  
In addition to providing a correct perturbative treatment of interparticle attractions 
they also divised the first prescription for mapping a softly repulsive reference 
system (required to treat e.g. Lennard-Jones particles) onto a system of hard-spheres 
with an effective, temperature-dependent diameter.  
The theory worked very well for a variety of model systems, accurately reproducing data 
for the thermodynamics and structure obtained from Monte-Carlo simulation. 
Although there nowadays exist more elaborate approaches to the thermodynamics, 
namely the self-consistent Ornstein-Zernike approximation of H\o ye and Stell 
\cite{scoza,stell} 
and the heirarchical reference theory of Reatto and Parola \cite{hrt}, these 
`beyond mean-field' approximations 
are not easy to implement and only yield significant differences from the Barker-Henderson 
theory in the vicinity of  the critical point. 

All of the aforementioned approaches have focused exclusively on homogeneous bulk states, 
for which the density is a constant. The treatment of fluids subject to external fields 
is much more difficult. 
A formal generalization of the bulk Barker-Henderson theory to inhomogeneous states is quite straightforward 
and leads naturally 
to an elegant density functional theory (see e.g.~\cite{Hansen06,evans79,evans92}). 
However, the implementation of this generalization necessitates calculation of  
inhomogeneous pair correlation functions and has thus never been 
seriously investigated. 
Although a number of simplified theories have been proposed 
\cite{averagerho1,averagerho2,averagerho3,averagerho4,averagerho5,averagerho6} 
they all rely on the dubious assumption 
that the inhomogeneous pair correlations of the 
reference system, which contain a great deal of subtle structural information,  
can be approximated by bulk pair correlation functions evaluated at an effective density. 
These empirical approaches have generally been used to study the free interface 
(a convenient test-case for which the density variation is smooth) but fail completely for 
strongly inhomogeneous systems.
Consequently, a quantitatively reliable theory for inhomogeneous fluids with attractive interactions 
is still lacking.  

The established work-horse of the density functional literature 
is a simplified mean-field approach in which the pair correlations of the reference 
system are treated in a crude approximation \cite{chacko}. 
This standard mean-field theory has proved very useful for exploring the rich phenomenology of 
inhomogeneous fluids, is easy to implement and does not require any reference to bulk states. 
Nevertheless, the standard theory remains unsatisfactory for two reasons:
Firstly, the predictions are not quantitative, which makes difficult a detailed 
comparison of theoretical predictions with data from experiment or simulation; 
effective parameters have to be 
chosen if data-fitting is to  be attempted. 
Secondly, it is possible that some of the phenomena predicted by the 
standard theory, such as layering transitions at substrates or other subtle packing effects 
at interfaces, could change qualitatively by incorporating a more correct treatment of 
internal correlations. 
       
In this paper we develop an accurate density functional approximation for nonuniform 
fluids with attractive interparticle interactions; the true inhomogeneous generalization 
of the Barker-Henderson theory.  
The paper will be structured as follows: 
In section \ref{theory} we will develop the theory, starting with the general equations 
for an arbitrary external field and then for the special cases of spherical and planar symmetry. 
In section \ref{results} we will focus on the hard-core Yukawa model and present numerical results 
for the density about a fixed test particle, between two confining walls and at the free interface. 
Finally, in section \ref{discussion} we will discuss the significance of our findings and 
provide an outlook for future work.

\section{Theory}\label{theory}

\subsection*{Classical density functional theory}
The density functional theory (DFT) provides an exact framework for the study of 
classical many-body systems under the influence of external fields 
\cite{evans79,evans92}. 
The central object of this approach is the grand potential functional
\begin{align}\label{grand}
\Omega[\,\rho\,] = F^{\rm id}[\,\rho\,] + F^{\rm exc}[\,\rho\,] 
- \int \!d\rv \big( \mu - V_{\rm ext}(\rv) \big)\rho(\rv), 
\end{align}
where $\mu$ is the chemical potential, $V_{\rm ext}(\rv)$ is the external potential, 
$\rho(\rv)$ is the one-body ensemble averaged density, and  
the square brackets indicate a functional dependence.
The Helmholtz free energy of the ideal gas is given by	
\begin{eqnarray}\label{idealfree}
F^{\rm id}[\,\rho\,]=k_BT\int\!d\rv\, \rho(\rv)\left(\, \ln(\rho(\rv))-1\, \right), 
\end{eqnarray}
where $k_B$ is the Boltzmann constant, $T$ is the temperature and 
we have set the thermal wavelength equal to unity. 
The excess Helmholtz free energy, $F^{\rm exc}[\,\rho\,]$, encodes the interparticle 
interactions and usually has to be approximated.
The grand potential satisfies the variational condition
\begin{align}
\label{EQomegaMinimial}
\frac{\delta  \Omega[\rho\,]}{\delta \rho(\rv)}=0, 
\end{align}
which generates an Euler-Lagrange equation for the equilibrium one-body density.

\subsection*{Exact free energy}
Although the excess free energy is not known in general, 
approximations can be facilitated by reexpressing it in terms of
the two-body density, $\rhotwo(\rv_1,\rv_2)$. 
This can be achieved by starting with the statistical mechanical result 
\cite{Hansen06,evans79,evans92} 
\begin{align}\label{pair_density}
\frac{\delta F}{\delta \phi(r_{12})}=\frac{1}{2}\,\rhotwo(\rv_1,\rv_2),
\end{align}
where $\phi(r_{12})\equiv\phi(|\rv_1-\rv_2|)$ is the full interaction potential 
and $F=F^{\rm id} + F^{\rm exc}$, and then formally integrating along 
a path in the function space of pair potentials. 
This operation, the inverse of functional differentiation, has been termed 
`functional line integration' \cite{line_integral} (see Appendix A). 
Application of this method to \eqref{pair_density} yields
\begin{align}\label{exact_MF}
F[\,\rho\,] = F_{\rm ref}[\,\rho\,] 
+ 
\frac{1}{2}\int_0^1 \!\!d\alpha \!\int\! d\rv_1\! \int\! d\rv_2\, 
\Delta\phi(r_{12})
\rhotwo_{\alpha}(\rv_1,\rv_2),
\end{align}
where we have split the full interaction potential into a sum of two terms, 
$\phi_{\alpha}= \phi_{\rm ref}+\alpha\Delta\phi$, where 
$\alpha$ is a `charging' parameter. 
If we define the difference $\Delta
\phi=\phi-\phi_{\rm ref}$ then 
increasing $\alpha$ from zero to unity 
enables us to go continuously from a reference system, characterized by interaction potential 
$\phi_{\rm ref}$, to the full system of interest. 
The first term on the r.h.s. of \eqref{exact_MF} is the Helmholtz free energy functional of 
the reference system (including the ideal gas contribution) and $\rhotwo_{\alpha}$ is the 
pair density of a system interacting via pair potential $\phi_{\alpha}$.

\subsection*{Perturbation approximation}

Equation \eqref{exact_MF} enables a clear mathematical expression of van der Waals' 
physical idea that liquid microstructure is dominated by interparticle repulsion. 
If we choose the repulsive part of the potential as a reference in \eqref{exact_MF}
and assume that the pair density does not change from 
that of the reference as $\alpha$ is turned on, then
we arrive at a perturbation theory for the free energy  of the fully interacting system. 
This is a mean-field approximation, because 
a pair density constructed using only the repulsive part of the interaction does not contain 
information about critical fluctuations. 
We will henceforth employ the hard-sphere system as our reference and split the full interaction 
potential into hard-sphere and attractive contributions, 
$\phi=\phi^{\rm hs} + \phi^{\rm att}$. 
Making the mean-field approximation leads directly to the Barker-Henderson (BH) 
functional
\begin{align}\label{inhom_bh}
F_{\rm BH}[\,\rho\,] &= F_{\rm hs}[\,\rho\,] 
\\
&\hspace*{-1.2cm}+ 
\frac{1}{2}\!\int\! d\rv_1\! \int\! d\rv_2\, 
\rho(\rv_1)\rho(\rv_2)\phi^{\rm att}(r_{12})
\big( 1 + h_{\rm hs}(\rv_1,\rv_2;[\,\rho\,]) \big),
\notag
\end{align}
where the first term is the free energy functional of the hard-sphere system 
(including the ideal gas contribution) and 
we have introduced the total correlation function 
\begin{align}\label{total_def}
h_{\rm hs}(\rv_1,\rv_2,[\,\rho\,]) = 
\frac{\rhotwo_{\rm hs}(\rv_1,\rv_2;[\,\rho\,])}{\rho(\rv_1)\rho(\rv_2)} - 1.   
\end{align}
%
The notation here has been chosen to make clear that for an inhomogeneous system 
the pair correlations are functionals of the one-body density. 
The density field thus enters \eqref{inhom_bh} both explicitly, via the quadratic density product 
in the integral, and implicitly, via the functional dependence of the reference free energy 
and reference total correlation function.

The primary difficulty in implementing \eqref{inhom_bh} is to find an accurate 
and tractable way to calculate $h_{\rm hs}(\rv_1,\rv_2;[\,\rho\,])$. 
The need to confront this issue, which is essentially the main point of the present work, can 
of course be avoided by simply setting the total correlation function equal to 
zero.     
This leads to the simplified expression 
\begin{align}\label{standard}
\hspace*{-0.2cm}
F_{\rm smf}[\,\rho\,] = F_{\rm hs}[\,\rho\,] 
+ 
\frac{1}{2}\!\int\! d\rv_1\!\! \int\! d\rv_2\, 
\rho(\rv_1)\rho(\rv_2)\phi^{\rm att}(r_{12}), 
\end{align}
the standard mean-field (SMF) functional \cite{sullivan}. 
This approximation has been used to investigate a variety of interfacial phenomena 
and, provided the reference hard-sphere functional is sufficiently 
accurate, does capture essential physical features \cite{evans79,evans92}.  
However, given the simplicity of the approximation, it is not surprising that 
the thermodynamic quantities obtained from the bulk limit of the SMF 
functional are in poor quantitative agreement with simulation data. 
There is also ambiguity regarding the 
definition of the attractive potential inside the region of hard-core repulsion, 
$\phi^{\rm att}(r_{12}<1)$; a feature which has been exploited, perhaps somewhat artificially, 
to introduce additional optimizing variational parameters \cite{orpa}. 

\subsection*{Bulk limit}
The bulk limit of the BH functional \eqref{inhom_bh} yields the following 
free energy density \cite{bh_original1,bh_original2,bh_review} 
\begin{align}\label{bulk_fe}
f_{\rm BH} = f_{\rm hs}
\;+\; \frac{1}{2}\,\rho_{\rm b}^2\!\int d\rv\; \phi^{\rm att}(r)\,
\left(1+h^{\rm b}_{\rm hs}(r)\right), 
\end{align}
where $\rho_{\rm b}$ is the bulk density and $h^{\rm b}_{\rm hs}$ is the 
bulk total correlation function.
Equation \eqref{bulk_fe} is the bulk free-energy of the first-order BH 
perturbation theory with a hard-sphere reference system. 
In their original work, Barker and Henderson also addressed 
softly repulsive reference systems 
by defining an effective sphere diameter and, moreover, suggested 
approximate forms for the second order term in the expansion \cite{bh_review,Hansen06}.

If the system phase separates, then the coexisting densities can be determined by requiring equality 
of the pressure and chemical potential in the two phases. 
The pressure is given by 
\begin{align}
P_{\rm BH} &= P_{\rm id} + P_{\rm hs} 
\\
&+ \frac{\rho_{\rm b}^2}{2}\!\!\int \!d\rv\, 
\phi^{\rm att}(r)\!\left(
1 \!+\! h^{\rm b}_{\rm hs}(r) \!+\! \rho_{\rm b}\frac{\partial h^{\rm b}_{\rm hs}(r)}{\partial \rho_{\rm b}}
\right),
\notag
\end{align}
where $P_{\rm id}\!=\!k_{\rm B}T\rho_b$ is the ideal contribution and 
$P_{\rm hs}$ is the excess pressure of the hard-sphere reference system. 
The van der Waals form for the equation of state is recovered only 
if the density dependence of $h^{\rm b}_{\rm hs}(r)$ is neglected, as would be the case for the 
standard mean-field theory. 
The chemical potential can be split into several terms,  
$\mu_{\rm BH} = \mu_{\rm id} +  \mu_{\rm hs} + \mu_{\rm smf} + \mu_{\rm corr} + \mu_{\rm der}$, 
where the ideal gas contribution is given by $\mu_{\rm id}=k_BT\ln(\rho_{\rm b})$ and 
those involving the attractive part of the interaction are given by 
\begin{align}
\mu_{\rm smf} &= \int d\rv\,\rho_{\rm b}\,\phi^{\rm att}(r),
\label{mu_smf}
\\
\mu_{\rm corr} &= \int d\rv\,\rho_{\rm b}\,\phi^{\rm att}(r)\,h^{\rm b}_{\rm hs}(r),
\label{mu_corr}
\\
\mu_{\rm der} &= 
\int d\rv\,\frac{\rho_{\rm b}^2\,\phi^{\rm att}(r)}{2}
\frac{\partial h^{\rm b}_{\rm hs}(r)}{\partial \rho_{\rm b}}.
\label{mu_deriv}
\end{align}
Within the well-known Percus-Yevick (PY) approximation \cite{py,Hansen06} there exist analytic expressions 
for both $h^{\rm b}_{\rm hs}$ and its density derivative \cite{smith,smith2} which  
facilitate accurate evaluation of the integrals in \eqref{mu_corr} and \eqref{mu_deriv}. 
The same PY approximation yields (via the compressibility route \cite{Hansen06}) the 
following expressions for the hard-sphere excess pressure 
\begin{align}
\beta P_{\rm hs} = \rho_b\left(
\frac{1 + \eta^2 + \eta^3}{(1-\eta)^3} - 1
\right)
\end{align}
and the hard-sphere excess chemical potential
\begin{align}
\beta\mu_{\rm hs} &= -\ln(1-\eta) + \frac{\eta}{(1-\eta)} + \frac{6\eta(1-\frac{3}{4}\eta)}{(1-\eta)^2} 
+ \frac{6\eta^2(1-\frac{1}{2}\eta)}{(1-\eta)^3} , 
\notag
\\
\label{mu_hs}
\end{align}
where $\beta=(k_BT)^{-1}$ and 
$\eta\!=\!\pi\rho_{\rm b}d^{\,3}/6$ is the packing fraction of hard-spheres with diameter $d$.

\subsection*{Euler-Lagrange equation}
We next consider implementation of the variational condition \eqref{EQomegaMinimial} 
specifically for the case of the BH functional. 
For the reference free energy we choose to employ the geometrically-based 
Rosenfeld functional for hard-spheres \cite{rosenfeld89}. Although several 
modified/improved variations of this functional have been proposed \cite{roth},
the original Rosenfeld formulation is sufficient to treat the situations 
to be considered in the present work 
(see Appendix B for details). 
Substituting equations \eqref{grand}, \eqref{idealfree} and \eqref{inhom_bh} into 
equation \eqref{EQomegaMinimial} generates the following Euler-Lagrange equation
\begin{align}\label{euler}
\rho(\rv) = e^{-\beta\left( \,V^{\rm ext}(\rv)\, -\, \mu \,-\, k_{\rm B}T c^{(1)}(\rv) \,\right) },
\end{align}
where we have set the thermal wavelength equal to unity.  
$c^{(1)}$ is the one-body direct correlation function, defined by the functional 
derivative 
\begin{align}\label{definition_of_c1}
c^{(1)}(\rv)=-\frac{\delta \beta F_{\rm BH}^{\rm exc}}{\delta\rho(\rv)}.
\end{align}
The quantity $-k_{\rm B}T c^{(1)}$ can be interpreted as an effective 
external field arising from interparticle interactions.  
Using \eqref{inhom_bh} to evaluate the derivative \eqref{definition_of_c1} generates 
four distinct contributions
\begin{align}\label{listoffour}
c^{(1)} = c_{\rm hs}^{(1)} + c_{\rm smf}^{(1)} + 
c_{\rm corr}^{(1)} + c_{\rm der}^{(1)},
\end{align}
where $c^{(1)}_{\rm hs}$ is the one-body direct correlation function of hard-spheres 
calculated from the Rosenfeld functional (see Appendix B). 
The remaining terms are given by 
\begin{align}
c_{\rm smf}^{(1)}(\rv_1) &=-\!\int\!\! d\rv_2\, \rho(\rv_2)\beta\phi^{\rm att}(r_{12}),
\label{c_smf}
\\
c_{\rm corr}^{(1)}(\rv_1) &=-\!\int\!\! d\rv_2\, \rho(\rv_2)\beta\phi^{\rm att}(r_{12})h_{\rm hs}(\rv_1,\rv_2),
\label{c_corr}
\\
c_{\rm der}^{(1)}(\rv_1)&=-\!\int\!\! d\rv_2\! \int\!\! d\rv_3 
\frac{\rho(\rv_2)\rho(\rv_3)\beta\phi^{\rm att}(r_{23})}{2 }\,
\label{c_deriv}
\frac{\delta h_{\rm hs}(\rv_2,\rv_3)}{\delta\rho(\rv_1)}.
\end{align} 
In order to solve the Euler-Lagrange equation \eqref{euler} 
we thus require a method to calculate both the two-body total correlation 
function appearing in \eqref{c_corr} and the functional derivative appearing in 
\eqref{c_deriv} - an intimidating three-body function. 
Although obtaining the latter quantity as a functional of the density 
is a difficult task, we will show that this is feasible in situations where
the external field has either planar or spherical symmetry.

\subsection*{Ornstein-Zernike equation}

The Ornstein-Zernike (OZ) equation for inhomogeneous fluids is an integral equation 
relating, for a given density profile, the 
two-body direct correlation function, $c_{\rm hs}$, to  the total correlation function 
\begin{align}\label{oz}
h^{}_{\rm hs}(\rv_1,\rv_2) = c_{\rm hs}^{}(\rv_1,\rv_2) \!+\! 
\int\! d\rv_3 h^{}_{\rm hs}(\rv_1,\rv_3)\rho(\rv_3) c_{\rm hs}^{}(\rv_3,\rv_2). 
\end{align} 
Although this equation applies for arbitrary interaction potential we will apply 
it only to the hard-sphere reference system, hence the subscript.
The external potential does not appear explicitly in this equation, but implicitly 
via its influence on the density. 
Equation \eqref{oz} can be regarded as the two-body analog of equation \eqref{euler} and 
serves to define $c^{}_{\rm hs}$ in terms of the density and total correlation function. 
Alternatively, $c^{}_{\rm hs}$ can be identified as the 
(negative) second functional derivative of the excess free energy
\begin{align}\label{derivative_def}
c^{}_{\rm hs}(\rv_1,\rv_2)=-\frac{\delta^2 
\beta F_{\rm hs}^{\rm exc}}{\delta\rho(\rv_1)\delta\rho(\rv_2)}.
\end{align}
There are thus two distinct paths by which the OZ equation can be used to obtain the total 
correlation function: (i) Given an approximation to the excess free energy functional, 
evaluate the second derivative \eqref{derivative_def} for the density of interest, substitute into 
\eqref{oz} and then solve for $h^{}_{\rm hs}$. 
(ii) Supplement \eqref{oz} by a second `closure' relation between 
$c_{\rm hs}^{}$ and $h^{}_{\rm hs}$, then 
solve self-consistently the two coupled equations. In the unlikely case that both the excess 
free energy and the closure relation are known exactly, then the two paths are equivalent. 	 

We are now faced with a choice of how best to calculate the total 
correlation function of the reference system, given that we are forced to use an 
approximate excess free energy functional when treating three-dimensional systems. 
The Rosenfeld functional is known to generate 
in most cases an accurate one-body direct correlation, as well as reliable {\it bulk} pair 
correlations when input to \eqref{derivative_def} followed by taking the homogeneous limit 
(the so-called OZ-route). 
However, the accuracy of the {\it inhomogeneous} pair correlations obtained from two functional 
derivatives of the Rosenfeld functional, particularly in situations for which the density 
is strongly varying, is less certain and remains to be systematically investigated. 
Taking path (i), described above, therefore risks conflation of error 
in the pair correlations of the reference system with the error inherent in a 
perturbative BH treatment of the attractive interaction. 
To make a clean assessment of the latter we are obliged to treat the 
reference system as accurately as possible and for this reason we will follow path 
(ii) to the pair correlations.  

A closure of the OZ equation which is known to be accurate for hard-spheres is the 
inhomogeneous Percus-Yevick approximation \cite{attard_book,attard1}
\begin{align}\label{py}
h^{}_{\rm hs}(\rv_1,\rv_2) &= -1 \;\;\;\text{for}\;  |\rv_1-\rv_2|<d,
\notag\\
c_{\rm hs}^{}(\rv_1,\rv_2) &= 0 \;\;\;\;\;\;\text{for}\; |\rv_1-\rv_2|>d.
\end{align} 
The first of these relations, the exact `core condition', expresses the impossibility  
of hard-sphere overlap, whereas the condition on $c_{\rm hs}$ is an approximation.  
While the PY theory has long been employed for studies of bulk 
fluids \cite{Hansen06} its inhomogeneous generalization is more rarely encountered. 
Numerical solution of equation \eqref{oz} for hard-spheres in the PY approximation can be 
facilitated using the simple rearrangement of the OZ equation outlined in Appendix C.

\subsection*{Three-body correlation function}

The most demanding task when implementing the Euler-Lagrange equation \eqref{euler} 
is the evaluation of the one-body direct correlation function contribution given by 
\eqref{c_deriv}. 
This requires the functional derivative of the total 
correlation function with respect to the density. 	
The first step in evaluating this quantity is to realize that the self-consistent 
solution of the coupled equations \eqref{oz} and \eqref{py} generates both 
the total and the two-body direct correlations as (implicit) functionals of the density.  
Given this observation the most straightforward way to calculate the functional derivative 
is to employ the physicists definition 
\begin{align}
\hspace*{-0.12cm}\frac{\delta h_{\rm hs}(\rv_1,\rv_2;[\,\rho\,])}{\delta\rho(\rv)} 
&=
\lim_{\varepsilon\rightarrow 0}
\frac{h^{}_{\rm hs}(\rv_1,\rv_2;[\,\rho^{}_\rv]) 
- h_{\rm hs}(\rv_1,\rv_2;[\,\rho\,])}{\varepsilon},
\notag
\\
&\equiv
\lim_{\varepsilon\rightarrow 0}
\frac{h^{\,\varepsilon\rv}_{\rm hs}(\rv_1,\rv_2) 
- h_{\rm hs}(\rv_1,\rv_2)}{\varepsilon}.
\label{derivative}
\end{align}
In the first equality we make explicit the functional dependence of the total correlation function 
on the density. 
If we choose the label $\rv_3$ as a dummy variable,
then $\rho^{}_\rv(\rv_3)=\rho(\rv_3) + \varepsilon\,\delta(\rv_3 - \rv)$
is the density as a function of $\rv_3$ subject to a local perturbation 
of amplitude $\varepsilon$ at the point $\rv$. 
The functional $h^{}_{\rm hs}(\rv_1,\rv_2;[\,\rho^{}_\rv])$ will then in general depend upon the three 
vector coordinates $\rv_1, \rv_2$ and $\rv$, where the latter can be viewed as an external parameter.
In the second equality of \eqref{derivative} we modify notation for later convenience, 
$h^{\,\varepsilon\rv}_{\rm hs}(\rv_1,\rv_2)$ being the hard-sphere total correlation function corresponding to 
the density perturbed at the point $\rv$. 
Using the definition \eqref{derivative} conveniently allows us to rewrite the derivative 
contribution to the one-body direct correlation function in the following simplified form
\begin{align}
c_{\rm der}^{(1)}(\rv_1)&= -\int d\rv_2\, 
\rho(\rv_2)
K(\rv_1,\rv_2),
\label{c_deriv_new}
\end{align} 
where the kernel is given by
\begin{align}\label{kernel}
K(\rv_1,\rv_2)=&
\\
&\hspace*{-1.8cm}\lim_{\varepsilon\rightarrow 0}\;
\frac{1}{2\,\varepsilon}
\Bigg(
c_{\rm corr}^{(1)}(\rv_2) 
+\! 
\int d\rv_3\, 
\rho(\rv_3)\beta\phi^{\rm att}(r_{23})\,
h^{\,\varepsilon\rv_1}_{\rm hs}(\rv_2,\rv_3)
\Bigg).
\notag
\end{align} 
The benefit of this rewriting is that the second term has the same 
structure (up to a parametric dependence on the coordinate $\rv_1$) as 
equation \eqref{c_corr} and so similar computer 
code can be used to evaluate both $c_{\rm corr}^{(1)}$ and 
$c_{\rm der}^{(1)}$. 
The final step is the determination of the perturbed total correlation function. 
Recalling that equations \eqref{oz} 
and \eqref{py} provide a functional map from the density to the pair 
correlations we substitute the perturbed density, $\rho_{\rv}$, into the OZ 
relation \eqref{oz}. This yields an integral equation for the perturbed 
total and direct correlation functions
\begin{align}\label{oz_epsilon}
h^{\varepsilon\rv}_{\rm hs}(\rv_1,\rv_2) &= c_{\rm hs}^{\,\varepsilon\rv}(\rv_1,\rv_2) 
+ \varepsilon\,h^{\,\varepsilon\rv}_{\rm hs}(\rv_1,\rv)c_{\rm hs}^{\,\varepsilon\rv}(\rv,\rv_2)
\notag\\
&+ \int\! d\rv_3\, h^{\,\varepsilon\rv}_{\rm hs}(\rv_1,\rv_3)\rho(\rv_3) 
c_{\rm hs}^{\,\varepsilon\rv}(\rv_3,\rv_2).   
\end{align}
This differs from the original OZ equation \eqref{oz} due to the second term. 
Equation \eqref{oz_epsilon} is closed by applying the PY conditions \eqref{py} to 
$h^{\varepsilon\rv}_{\rm hs}$ and $c^{\varepsilon\rv}_{\rm hs}$ and iterating to 
convergence for fixed $\varepsilon$ and $\rv$.

\subsection*{Spherical geometry} 

Now that we have the relevant equations in their general form we will consider the 
special case where the density has spherical symmetry.
This enables the integrals occuring 
in \eqref{oz} and \eqref{oz_epsilon} to be reduced exactly to one-dimension, 
greatly facilitating their numerical evaluation.    
The appropriate method is expansion in Legendre polynomials. 
A spherically inhomogeneous two-body function requires as input three independent 
variables; two radial distances and the angle 
between them. 
For example, the total correlation function 
\begin{align}\label{h_spherical}
h_{\rm hs}(\rv_1,\rv_2)\rightarrow h_{\rm hs}^{\rm sp}(r_1,r_2,x_{12}), 
\end{align}
where $x_{12}=\cos(\theta_{12})$. 
The Legendre transform of a spherically inhomogeneous two-body function 
is given by
\begin{align}\label{leg_transform}
H_n(r_1,r_2)=\frac{2n+1}{2}\int_{-1}^{+1}\!dx_{12}\;h^{\rm sp}_{\rm hs}(r_1,r_2,x_{12})\,P_n(x_{12}), 
\end{align}
where $P_n(x)$ is a Legendre polynomial. 
Numerical evaluation of \eqref{leg_transform} requires a discretization scheme capable 
of handling the highly oscillatory structure of the higher-order Legendre polynomials. 
We thus use the Gauss-Legendre quadrature proposed by Attard \cite{attard1}. 
The back-transform is given by
\begin{align}\label{reconstruct}
h^{\rm sp}_{\rm hs}(r_1,r_2,x_{12})=\sum_{n=0}^{\infty} H_n(r_1,r_2)\,P_n(x_{12}). 
\end{align}
In practice the sum can be truncated at a finite number of terms, depending on the level 
of accuracy required. 
Taking the Legendre transform of the OZ equation \eqref{oz} reduces the three-dimensional integral 
to a radial integral
\begin{align}\label{oz_sph}
H_n(r_1,r_2) &= C_n(r_1,r_2)
\\ 
&+ \frac{4\pi}{2n+1}\int_0^{\infty}dr_3\,r_3^2\,H_n(r_1,r_3)\rho(r_3)\,C_n(r_3,r_2). 
\notag
\end{align}
Determination of the pair correlations $h_{\rm hs}^{\rm sp}$ 
and $c_{\rm hs}^{\rm sp}$ proceeds by iterating between 
\eqref{oz_sph} and the PY closure \eqref{py}. 
For hard-spheres special care has to be taken to accurately transform the 
discontinuous pair correlations. An accurate method to deal with this problem 
is described in the Appendix of Ref.\cite{attard1}. 
Once $h^{\rm sp}_{\rm hs}$ has been determined we can evaluate the 
correlation contribution 
\begin{align}
c_{\rm corr}^{(1)}(r_1) &=-4\pi\!
\int_0^{\infty} \!dr_2 \,r_2^2\,
\rho(r_2)\,U(r_1,r_2),
\end{align}
where $U$ is the $n=0$ Legendre transform of the product of the reduced 
interaction potential with the total correlation function,
\begin{align}
U(r_1,r_2)=\frac{1}{2}\int_{-1}^{+1}\!dx_{12}\;
\beta\phi^{\rm att}(r_{12})\,h^{\rm sp}_{\rm hs}(r_1,r_2,x_{12}),
\end{align}
and we recall that $r^2_{12}=r_1^2 + r_2^2 - 2r_1r_2x_{12}$\,. 

Evaluation of the remaining contribution to the one-body direct correlation function 
\eqref{c_deriv} requires careful handling of functional derivatives 
in the spherical coordinate system. 
Consideration of the dimensionality and radial scaling of the functional 
derivative leads to
\begin{align}
\frac{\delta h_{\rm hs}(\rv_1,\rv_2)}{\delta\rho(\rv)}
=\frac{1}{4\pi r^2}\frac{\delta h^{\rm sp}_{\rm hs}(r_1,r_2,x_{12})}{\delta\rho(r)}.
\end{align}
If we again employ the physicists finite difference definition then we obtain 
\begin{align}\label{spherical_rewrite}
\frac{\delta h_{\rm hs}(\rv_1,\rv_2)}{\delta\rho(\rv)}
=
\lim_{\varepsilon\rightarrow 0}
\frac{
h^{\rm sp,\varepsilon r}_{\rm hs}(r_1,r_2,x_{12})
- 
h^{\rm sp}_{\rm hs}(r_1,r_2,x_{12})
}{4\pi r^2\varepsilon},
\end{align}
where $h^{\rm sp,\varepsilon r}_{\rm hs}$ is the total correlation function corresponding to 
the perturbed density $\rho^{}_r(r_3)=\rho(r_3) + \varepsilon\,\delta(r_3 - r)$. 
Equation \eqref{c_deriv_new} thus becomes
\begin{align}\label{sp_deriv}
c_{\rm der}^{(1)}(r_1)&= -\int_{0}^{\infty} \!dr_2 \left(\frac{r_2}{r_1}\right)^2 
\rho(r_2)
K_{\rm sp}(r_1,r_2),
\end{align} 
where the kernel is given by
\begin{align}
&K_{\rm sp}(r_1,r_2)=
\\
&\lim_{\varepsilon\rightarrow 0}\;
\frac{1}{2\,\varepsilon}\Bigg(
c_{\rm corr}^{(1)}(r_2) 
+
4\pi\!
\int_0^{\infty} \!dr_3 \,r_3^2\,
\rho(r_3)\,U_{\rm sp}^{\varepsilon r_1}(r_2,r_3)
\Bigg),
\notag
\end{align} 
and $U_{\rm sp}^{\varepsilon r_1}$ is given by
\begin{align}
U_{\rm sp}^{\varepsilon r_1}(r_2,r_3)=\frac{1}{2}\int_{-1}^{+1}\!dx_{23}\;
\beta\phi^{\rm att}(r_{23})\,h^{\rm sp,\varepsilon r_1}_{\rm hs}(r_2,r_3,x_{23}).
\end{align}
It remains to find an equation to determine $h^{\rm sp,\varepsilon r}_{\rm hs}$. 
Substitution of the perturbed density, $\rho_r$, into the transformed equation 
\eqref{oz_sph} yields 
\begin{align}\label{oz_sph_pert}
H^{\varepsilon r}_n(r_1,r_2) &= C^{\varepsilon r}_n(r_1,r_2)
+ \frac{4\pi r^2\varepsilon}{2n+1} H^{\varepsilon r}_n(r_1,r)\,C^{\varepsilon r}_n(r,r_2)
\notag\\ 
&\hspace*{-0.5cm}+ \frac{4\pi}{2n+1}\int_0^{\infty}\!dr_3 \,r_3^{2}
\,H^{\varepsilon r}_n(r_1,r_3)\rho(r_3)\,C^{\varepsilon r}_n(r_3,r_2), 
\end{align}
where the Legendre transformed pair correlation functions have a parametric 
dependence on the amplitude and position of the density perturbation.
Equation \eqref{oz_sph_pert} has to be solved together with the PY closure \eqref{py} for all 
required values of the coordinate $r$. 

Some points which are important for an 
efficient computational implementation:   
(i) The fact that $r$ enters here as an external parameter allows the solution 
of the coupled equations \eqref{py} and \eqref{oz_sph_pert} to be performed in 
parallel for different values of $r$. 
(ii) Once $H^{\varepsilon r}_n(r_1,r_2)$ is known for a given value of the external 
coordinate $r$ we can use it to evaluate $c^{(1)}_{\rm der}$ and 
then discard $H^{\varepsilon r}_n(r_1,r_2)$. The storage of a large array can 
thus be avoided. 
(iii) The symmetry of the pair correlations can be exploited, for example 
the invariance of the total correlation function with respect to exchange of arguments 
implies that $H^{\varepsilon r}_n(r_1,r_2)=H^{\varepsilon r}_n(r_2,r_1)$.

\subsection*{Planar geometry}

The second special case of interest is that of planar symmetry, for which 
the density only varies as a function of a single cartesian coordinate (we choose the 
$z$-axis). 
The inhomogeneous pair correlations exhibit cylindrical symmetry and depend upon 
two coordinates and a cylindrical radial distance separating them
\begin{align}
h_{\rm hs}(\rv_1,\rv_2)\rightarrow h_{\rm hs}^{\rm pl}(z_1,z_2,\bar{r}_{12}). 
\end{align}
The direct separation $r_{12}$ between two points in space, $\rv_1$ and $\rv_2$, 
is related to the cylindrical separation $\bar{r}_{12}$ according to 
$r_{12}^{2}=(z_1-z_2)^2 + \bar{r}_{12}^2$. 
The appropriate method to apply in this case is the Hankel transform  
\begin{align}
\mathcal{H}_{k}(z_1,z_2)
= 
2\pi\int_0^{\infty} \!d\bar{r}_{12}\, \bar{r}_{12}J_0(k\bar{r}_{12})
h^{\rm pl}_{\rm hs}(z_1,z_2,\bar{r}_{12}),
\end{align}
which is simply a two-dimensional Fourier transform in the plane perpendicular to 
the $z$-axis.
$J_0$ is the zeroth-order Bessel function of the first kind. 
The inverse Hankel transformation is given by
\begin{align}
h^{\rm pl}_{\rm hs}(z_1,z_2,\bar{r}_{12}) 
= 
\frac{1}{2\pi}\int_0^{\infty} \!dk\, kJ_0(k\bar{r}_{12})
\mathcal{H}_{k}(z_1,z_2).
\end{align}
For our numerical calculations we employ the efficient and accurate discretization 
scheme of Lado \cite{lado}.
Application of the Hankel transform to the OZ equation \eqref{oz} leads to the simplified form
\begin{align}\label{oz_planar}
\mathcal{H}_{k}(z_1,z_2) = \mathcal{C}_{k}(z_1,z_2)
+\! \int_{-\infty}^{\infty} \!\!\!\!dz_3 \, 
\mathcal{H}_{k}(z_1,z_3)
\rho(z_3)
\mathcal{C}_{k}(z_3,z_2),
\end{align}
where $\mathcal{C}_{k}$ is the Hankel transform of the direct correlation function.
Unlike the case of spherical geometry, the correct way to treat the discontinuous pair 
correlation functions has not previously been documented and we thus direct the reader 
to Appendix C for details. 
The correlation contribution to the one-body direct correlation function is given by
\begin{align}\label{c1_corr_pl}
c^{(1)}_{\rm corr}(z_1)=-\int_{-\infty}^{\infty}dz_2 \,\rho(z_2)\,W(z_1,z_2),
\end{align}
where $W$ is the zero wavevector Hankel transform of the product of the reduced interaction 
potential with the total correlation function 
\begin{align}
W(z_1,z_2) = 2\pi\int_{0}^{\infty}d\bar{r}_{12}\, \bar{r}_{12}\, \beta\phi^{\rm att}(r_{12})
\,h_{\rm hs}^{\rm pl}(z_1,z_2,\bar{r}_{12}).
\end{align}
The functional derivative required for evaluation of \eqref{c_deriv} 
can be reexpressed in terms of a derivative with respect to the one-dimensional density 
profile
\begin{align}
\frac{\delta h_{\rm hs}(\rv_1,\rv_2)}{\delta\rho(\rv)}
=\frac{1}{A}\frac{\delta h^{\rm pl}_{\rm hs}(z_1,z_2,\bar{r}_{12})}{\delta\rho(z)},
\end{align}
where $A$ is an (arbitrary) area perpendicular to the $z$-axis which will 
cancel-out in subsequent calculations. 
Using finite differences the derivative becomes 
\begin{align}\label{planar_rewrite}
\frac{\delta h_{\rm hs}(\rv_1,\rv_2)}{\delta\rho(\rv)}
=
\lim_{\varepsilon\rightarrow 0}
\frac{
h^{\rm pl,\varepsilon z\!}_{\rm hs}(z_1,z_2,\bar{r}_{12})
- 
h^{\rm pl}_{\rm hs}(z_1,z_2,\bar{r}_{12})
}{A\varepsilon},
\end{align}
where $h^{\rm pl,\varepsilon z}_{\rm hs}$ is the total correlation function corresponding to 
the perturbed density $\rho^{}_z(z_3)=\rho(z_3) + \varepsilon\,\delta(z_3 - z)$. 
Equation \eqref{c_deriv_new} thus becomes
\begin{align}\label{pl_deriv}
c_{\rm der}^{(1)}(z_1)= -\int_{-\infty}^{\infty} \!dz_2 
\,\rho(z_2)
K_{\rm pl}(z_1,z_2),
\end{align} 
where the kernel is given by
\begin{align}
&K_{\rm pl}(z_1,z_2)=
\\
&\lim_{\varepsilon\rightarrow 0}\;
\frac{1}{2\,\varepsilon}\Bigg(
c_{\rm corr}^{(1)}(z_2) 
+
\int_{-\infty}^{\infty} \!dz_3\,
\rho(z_3)\,U_{\rm pl}^{\varepsilon z_1\!}(z_2,z_3)
\Bigg).
\notag
\end{align} 
The first term in this expression is known already from equation \eqref{c1_corr_pl} 
and $U_{\rm pl}^{\varepsilon z_1}$ is given by
\begin{align}
U_{\rm pl}^{\varepsilon z_1\!}(z_2,z_3)= 2\pi\!\int_{0}^{\infty}\!\!\!d\bar{r}_{23}\, 
\bar{r}_{23}\,
\beta\phi^{\rm att}(r_{23})\,h^{\rm pl,\varepsilon z_1\!}_{\rm hs}(z_2,z_3,\bar{r}_{23}).
\end{align}
The integral equation required to determine $h^{\rm pl,\varepsilon z_1}_{\rm hs}$ 
is obtained by substituting the perturbed density, $\rho_z$, into the transformed OZ 
equation \eqref{oz_planar}. This yields the following expression
\begin{align}\label{oz_planar_epsilon}
\hspace*{-0.17cm}\mathcal{H}^{\varepsilon z\!}_{k}(z_1,z_2) &= 
\mathcal{C}^{\varepsilon z\!}_{k}(z_1,z_2)
+
\varepsilon\,\mathcal{H}^{\varepsilon z\!}_{k}(z_1,z) 
\mathcal{C}^{\varepsilon z\!}_{k}(z,z_2)
\notag\\
&+\int_{-\infty}^{\infty} \!\!\!dz_3 \, 
\mathcal{H}^{\varepsilon z\!}_{k}(z_1,z_3)
\rho(z_3)
\mathcal{C}^{\varepsilon z\!}_{k}(z_3,z_2).
\end{align}
Equation \eqref{oz_planar_epsilon} is to be solved together with the PY closure \eqref{py} for all 
required values of the parameter $z$.

\subsection*{Numerical strategy and simulation details}

Our general numerical scheme for determining the density profile proceeds in the following way: 
\\

\noindent{\bf (i)} Select an initial guess for the density and evaluate all contributions to the one-body 
direct correlation function, see Eq.~\eqref{listoffour}. 
Evaluation of $c^{(1)}_{\rm corr}(\rv)$ and $c^{(1)}_{\rm der}(\rv)$ requires solution of 
the relevant inhomogeneous integral equations (Eqs.~\eqref{oz_sph} and \eqref{oz_sph_pert} in spherical geometry, 
Eqs.~\eqref{oz_planar} and \eqref{oz_planar_epsilon} in planar geometry), 
which we perform using a simple Picard iteration with Broyles 
mixing \cite{Hansen06}.
When possible check the bulk limits of the various contributions using 
\eqref{mu_smf}, \eqref{mu_corr} and \eqref{mu_deriv}. 
\\

\noindent{\bf (ii)} Keeping the functions $c^{(1)}_{\rm corr}(\rv)$ and $c^{(1)}_{\rm der}(\rv)$ 
fixed we iterate the Euler-Lagrange equation \eqref{euler} to convergence to obtain a 
new estimate for the density. Here we again employ simple Picard iteration. 
\\

\noindent{\bf (iii)} Update $c^{(1)}_{\rm corr}(\rv)$ and go back to step (ii). 
Keep iterating between steps (ii) and (iii) until both the density and $c^{(1)}_{\rm corr}(\rv)$ 
have converged. During this process the function $c^{(1)}_{\rm der}(\rv)$ is {\it not} modified. 
\\

\noindent{\bf (iv)} Update $c^{(1)}_{\rm der}(\rv)$ and return to step (ii). As this is the most 
computationally expensive step we aim to keep the number of these updates to a minimum 
(at most three to four iterations were required for the situations considered in this work). 
The process is terminated when both the direct correlation function contributions and the 
density have converged.
\\

This protocol provides reliable and stable convergence in all cases studied and 
avoids unneccessary function evaluations. 
However, we realize that this is only one of many possible schemes and may not be the most efficient 
strategy. 
It is also likely that computational time could be reduced using more sophisticated methods 
to solve the integral equations (e.g. conjugate gradient), but we have chosen to prioritize 
accuracy and stability over speed.

The simulation data were generated using standard methods \cite{Allen}. 
To calculate the radial distribution function we employed canonical Monte-Carlo (MC), 
with 432 particles and periodic boundary conditions. 
The potential was truncated at $r=3d$ (not shifted).
To calculate the density profiles in slit confinement we used grand canonical Monte-Carlo 
simulations (GCMC) with periodic boundary conditions in the $x$ and $y$ directions. 
The length of the box in these directions was $25d$ and the potential was truncated at $r=5d$ 
(not shifted). We have checked the robustness of our predictions with respect to these choices 
of numerical parameters.

\section{Results}\label{results}
For our numerical calculations we will consider the hard-core Yukawa (HCY) interaction potential
\begin{align}\label{hcy}
\phi^{\rm att}(r_{12}) = 
\begin{cases} 
      \hspace*{0.9cm}\infty & r_{12} < 1, \vspace*{0.2cm}\\
      -\kappa\,\frac{e^{-\alpha(r_{12}-1)}}{r_{12}} & r_{12}\ge 1,
\end{cases}
\end{align}
where $\kappa$ and $\alpha$ are positive constants. Here and in the following all 
lengths are measured in units of a hard-sphere diameter. 
For the remainder of this work we will focus on the well-studied special case $\alpha=1.8$, 
which is similar in range to the standard Lennard-Jones potential.

\subsection*{Bulk phase diagram}
\begin{figure}[!t]
\centering
\includegraphics[width=0.92\linewidth]{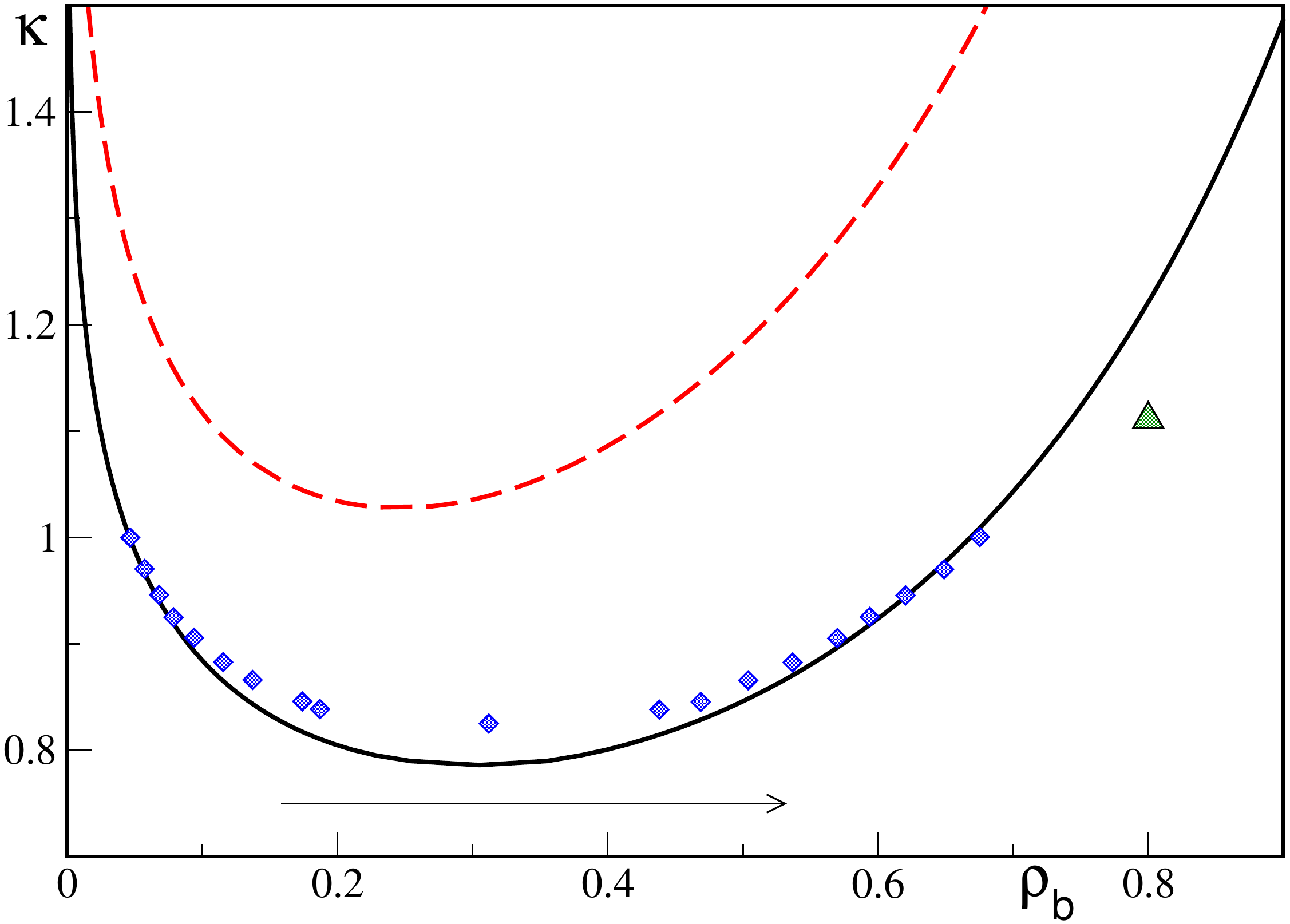}
\caption{
{\bf Phase diagram} for $\alpha=1.8$. 
Standard mean-field theory (broken red like), 
BH theory (full black line) and MC simulation data taken from 
Ref.\cite{stell}. 
The triangle indicates the state-point at which we show the radial distribution 
in Fig.\ref{fig:radial_highrho} and the arrow indicates the path taken when calculating 
the density profiles in Fig.\ref{BHslitprofiles}. 
}
\label{fig:phase}
\end{figure}

In Fig.\ref{fig:phase} we show the bulk phase boundary (binodal) from the SMF and BH theories alongside 
accurate MC simulation data taken from Ref.\cite{stell}. 
The simulation critical point is estimated to be at $\kappa_{\rm crit}\!\approx\!0.84$
and $\rho_{\rm crit}\!\approx\!0.3$.
The BH theory improves significantly upon the predictions of the SMF theory and accurately captures 
the values of the coexisting densities as $\kappa$ is increased towards the triple point, 
which we estimate to be at a density $\rho_{\rm tr}\!\approx\!0.9$ \cite{triple}. 
This trend is consistent with previous studies for Lennard-Jones and 
square-well fluids \cite{bh_review}. 
In the critical region we observe the expected discrepancies arising from the mean-field 
approximation; 
we can thus anticipate that inhomogeneous BH calculations will be the least reliable 
at state-points close to the bulk critical point. 

\subsection*{Test particle}
As a first test of the BH functional we will focus 
on a situation where the external field is a fluid particle fixed at the origin 
\begin{align}\label{Vext_testparticle}
V_{\rm ext}(r) = 
\begin{cases} 
      \hspace*{0.9cm}\infty & r < 1, \vspace*{0.2cm}\\
      -\kappa\,\frac{e^{-\alpha(r-1)}}{r} & r\ge 1.
\end{cases}
\end{align}
The significance of this choice is that the inhomogeneous density about a test particle 
is related to the bulk radial distribution function according to the Percus identity
$g(r)=\rho(r)/\rho_b$ \cite{evans92} and thus provides direct access to bulk thermodynamic quantities. 
Numerical minimization of the BH functional was performed on a discrete spatial grid with spacing 
$\Delta r=0.05$ and 
using 180 Legendre polynomials. 
We have checked carefully the robustness of the converged density profiles to variations in the 
choice of these numerical parameters.   


\begin{figure}[!t]
\centering
\includegraphics[width=0.95\linewidth]{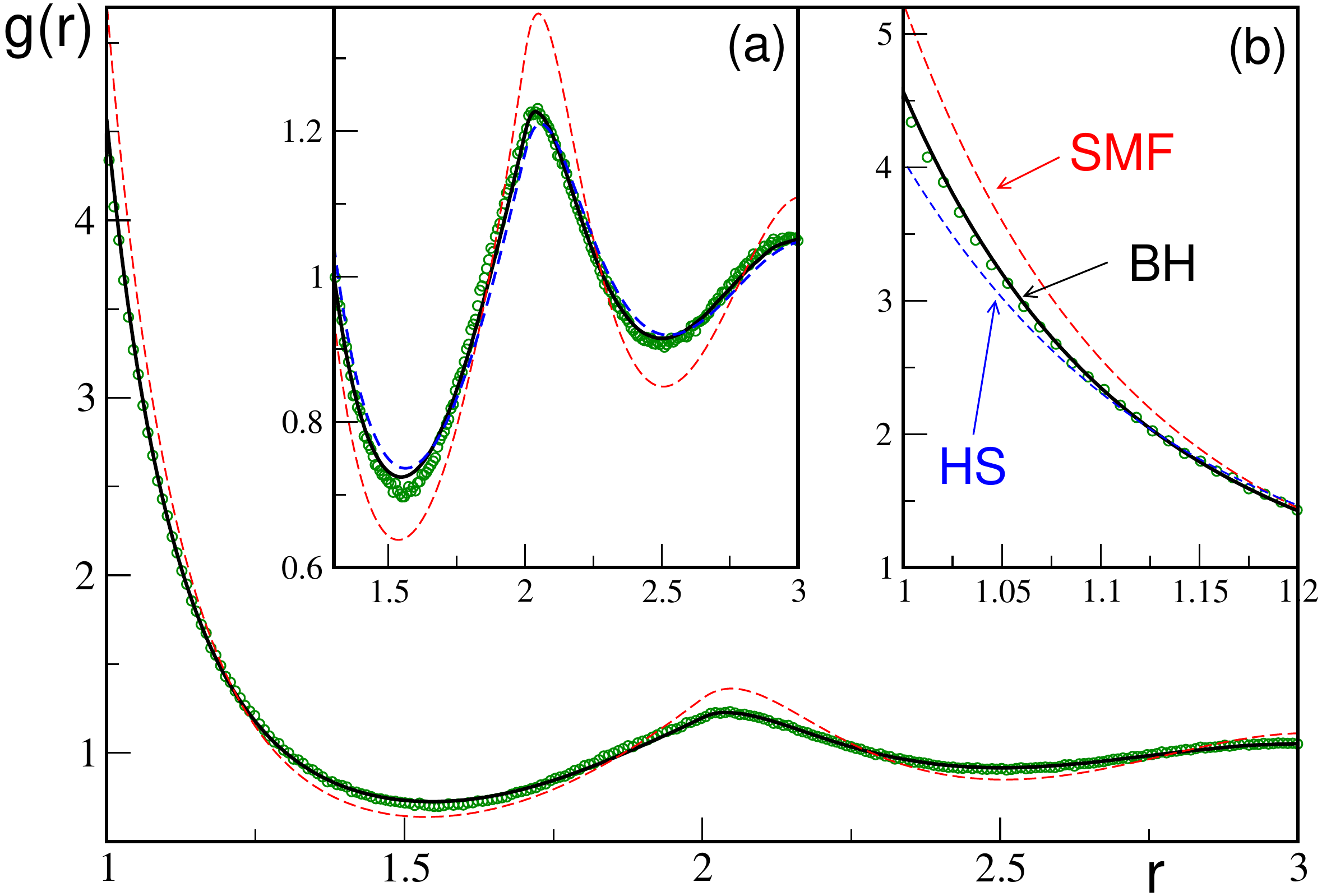}
\caption{
{\bf Test particle.} 
Comparison of the radial distribution function calculated using the test particle 
method with simulation data at $\kappa\!=\!1.111$ and $\rho_b=0.8$ (marked with a 
triangle in Fig.\ref{fig:phase}). 
Green circles: MC simulation.
Full black line: BH functional. 
Broken red line: SMF theory. 
Dashed blue line: The density of pure hard-spheres 
($\kappa\!=\!0$, $\rho_b\!=\!0.8$) calculated using the Rosenfeld functional. 
Insets (a) and (b) focus on the second peak and contact value, respectively.
}
\label{fig:radial_highrho}
\end{figure}

In Fig.\ref{fig:radial_highrho} we compare $g(r)$ calculated using the SMF and the BH theories with 
MC data for the statepoint at $\kappa=1.111$ and $\rho_b=0.8$ (indicated by the triangle in 
Fig.\ref{fig:phase}). 
We find that the SMF significantly overestimates the structure in $g(r)$ compared to the 
simulation. 
The first (contact) peak is around $16$\% too high and the amplitude of subsequent 
oscillations is too large. 
These features are consistent with the findings of Archer {\it et al.} \cite{chacko}, who 
assessed the performance of the SMF in one-dimensional test particle calculations using an 
exactly solvable model as a benchmark. 
The BH theory provides an accurate description of the simulation data, showing only 
small errors in the contact value and depth of the first minimum. 
It is interesting to note that, despite the large value of $\kappa$, the BH $g(r)$ is 
very similar to that of pure hard-spheres (also shown in Fig.\ref{fig:radial_highrho}).
This observation validates a posteriori the van der Waals picture that repulsive interactions
dictate the microstructure and is consistent with the perturbation approximation 
at the heart of BH theory.

\begin{figure}[!t]
\centering
\includegraphics[width=0.825\linewidth]{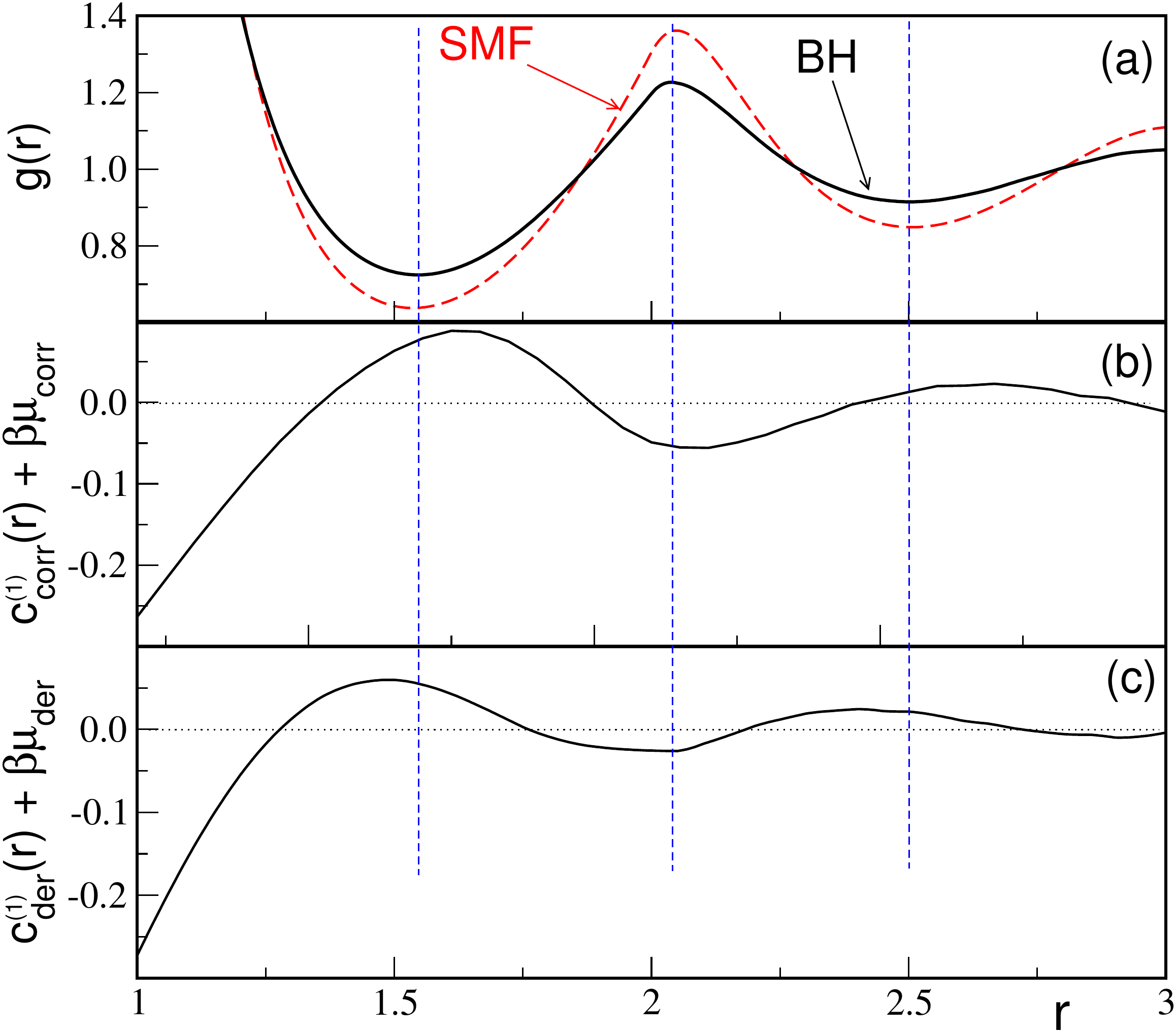}
\caption{
{\bf Test particle.} 
The companion to Fig.2 showing the individual contributions to the 
one-body direct correlation function appearing in equation \eqref{euler}. 
Full black lines: BH functional. 
Broken red line: SMF theory.
The blue lines are a guide for the eye to show how the maxima and minima of these 
functions match up with the oscillations in the radial distribution function.
}
\label{fig:direct_terms}
\end{figure}

On the level of the Euler-Lagrange equation \eqref{euler} the difference between the SMF and 
the BH theories is due to the direct correlation contributions 
$c^{(1)}_{\rm corr}$ and $c^{(1)}_{\rm der}$, which we show in Fig.\ref{fig:direct_terms}. 
Given the structural overestimation of the 
SMF theory, these self-consistently determined functions apparently serve 
to counteract the term $c^{(1)}_{\rm smf}$ and thus yield a radial distribution function 
very similar to that of hard-spheres. 
Although the oscillations in $c^{(1)}_{\rm corr}$ and $c^{(1)}_{\rm der}$ are not 
in phase with each other, the peaks and troughs act to suppress the exaggerated 
oscillations occuring in the SMF theory. 
We also observe that both contributions are of comparable magnitude; 
neglecting $c^{(1)}_{\rm der}$, which would be highly desirable from a computational standpoint, is 
therefore not a viable option.  

\subsection*{Planar slit}
We next consider the density of the HCY fluid confined between two hard-walls 
separated by a distance $L$ and oriented perpendicular to the $z$-axis. 
The external potential is given by
\begin{align}\label{Vext_walls}
V_{\rm ext}(z) = 
\begin{cases} 
      \hspace*{0.07cm}0& \frac{1}{2}<z<L-\frac{1}{2}, \vspace*{0.2cm}\\
       \infty  &\hspace*{0.25cm}{\rm otherwise},
\end{cases}
\end{align}
where we recall that the unit of length is taken to be one particle diameter. 
Numerical results will be presented for the case $L\!=\!10$.
The BH functional was minimized on a grid with spacing $\Delta z=0.05$. 
When employing the Lado discrete Hankel transform (see Ref.~\cite{lado} for details) 
it is neccessary 
to specify a cutoff length, $R$, in the plane parallel to the interface and 
a maximum number of radial grid points located at the zeros of the Bessel function 
$J_0$. We found that using $R\!=\!14$ and $200$ Bessel zeros provided very accurate 
results. We carefully checked that the converged density profiles were robust with 
respect to changes in the numerical parameters.

\begin{figure}
\hspace*{-0.3cm}
\includegraphics[width=0.95\linewidth]{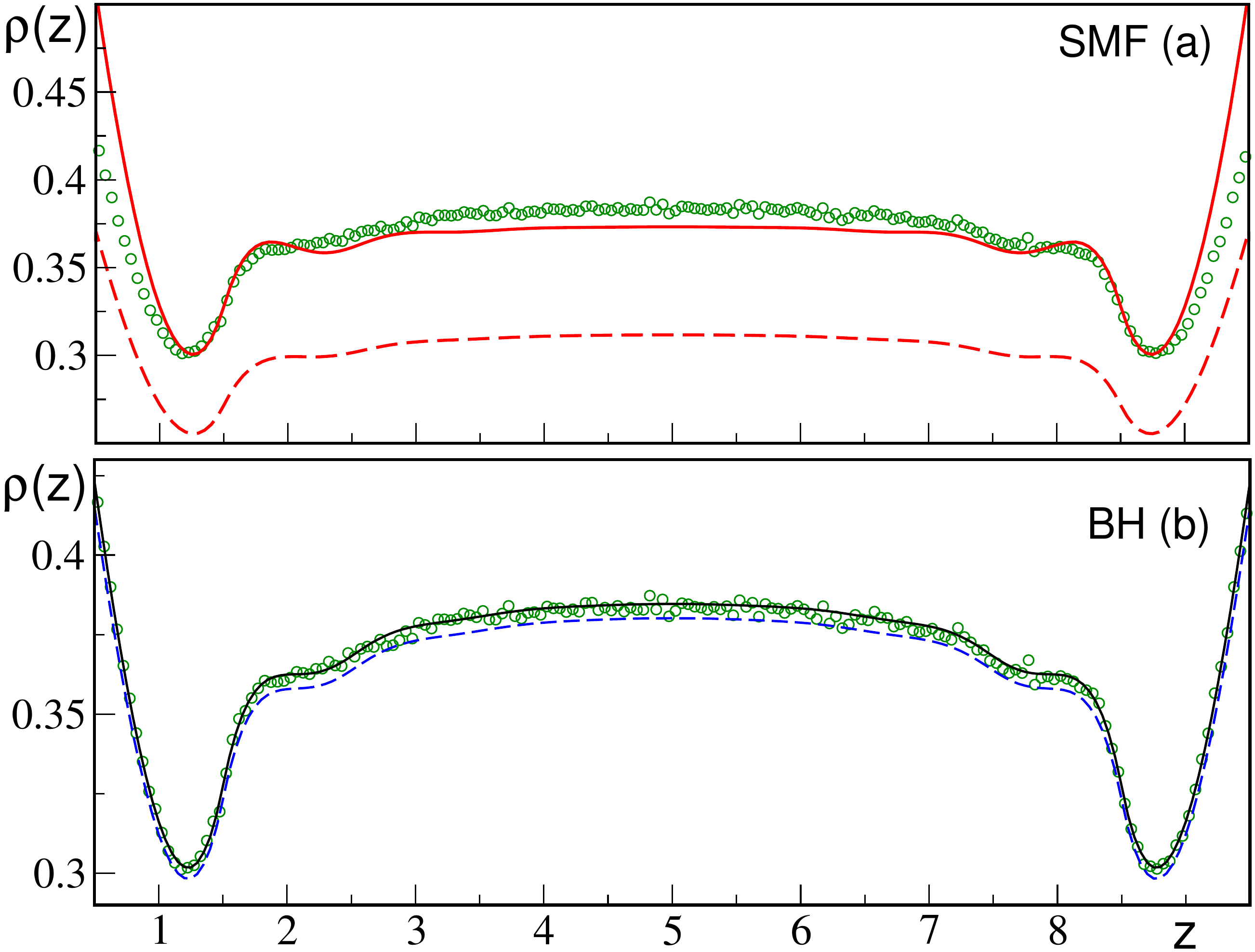}
\caption{
{\bf Planar slit.} The density between two hard-walls located at $z\!=\!0$ and $10$ 
for $\kappa\!=\!0.5$. Green circles: GCMC data at $\mu=-1$. 
DFT profiles calculated at $\mu=-1$ are given by the broken blue line (BH functional) 
and the broken red line (SMF functional).
DFT profiles calculated by adjusting $\mu$ such that 
the average number of particles in the system matches that of simulation are given by the 
 full black line (BH functional) and the full red line (SMF functional). 
}
\label{slit_easypoint}
\end{figure}

In Fig.\ref{slit_easypoint} we show GCMC data together with the density 
obtained from the SMF and BH theories, respectively. 
The theoretical results indicated by broken lines and the simulation data points
were calculated at $\kappa=0.5$ and chemical potential $\mu=-1$.    
As we have a confined system we now specify the chemical potential 
rather than a bulk density, as the latter is no longer well-defined. 
The BH functional captures the simulation data very well, only slightly 
underestimating the density at the centre of the gap. 
In contrast, the SMF functional underestimates this value by around $26$\% and provides a 
generally poor description of the simulation data. 

The situation discussed above, for which calculations are performed 
at the same chemical potential as the GCMC simulations, is the correct way to test the 
quality of an approximate DFT; first-principles predictions are made and then tested.  
An important factor in determining the form of the density profile is the 
proximity of the chosen state point to bulk phase coexistence.
However, if one wishes to use a given DFT approximation to fit existing simulation (or 
indeed experimental) data, then better results can be obtained by treating $\mu$ as an 
optimization parameter. 
The chemical potential can be tuned such that the average number of particles in the 
system (i.e.~the integral of the density profile) from theory matches that from 
simulation. 
The full curves shown in Fig.\ref{slit_easypoint} are the result of such a fitting 
procedure. 
For the BH functional the chemical potential need only be tuned away from the simulation 
value ($\mu=\!-1$) by around $1$\% to match the average particle number, 
resulting in a very close fit. 
In contrast, the SMF theory requires much more substantial adjustment of $\mu$ and, 
even then, the resulting fit is not satisfactory. We see here, consistent with 
Fig.\ref{fig:radial_highrho}, that the SMF theory tends to overestimate the structure of 
the density profile, particularly in regions close to a strongly repulsive boundary.

\begin{figure}[!t]
\hspace*{-0.9cm}
\includegraphics[width=0.9\linewidth]{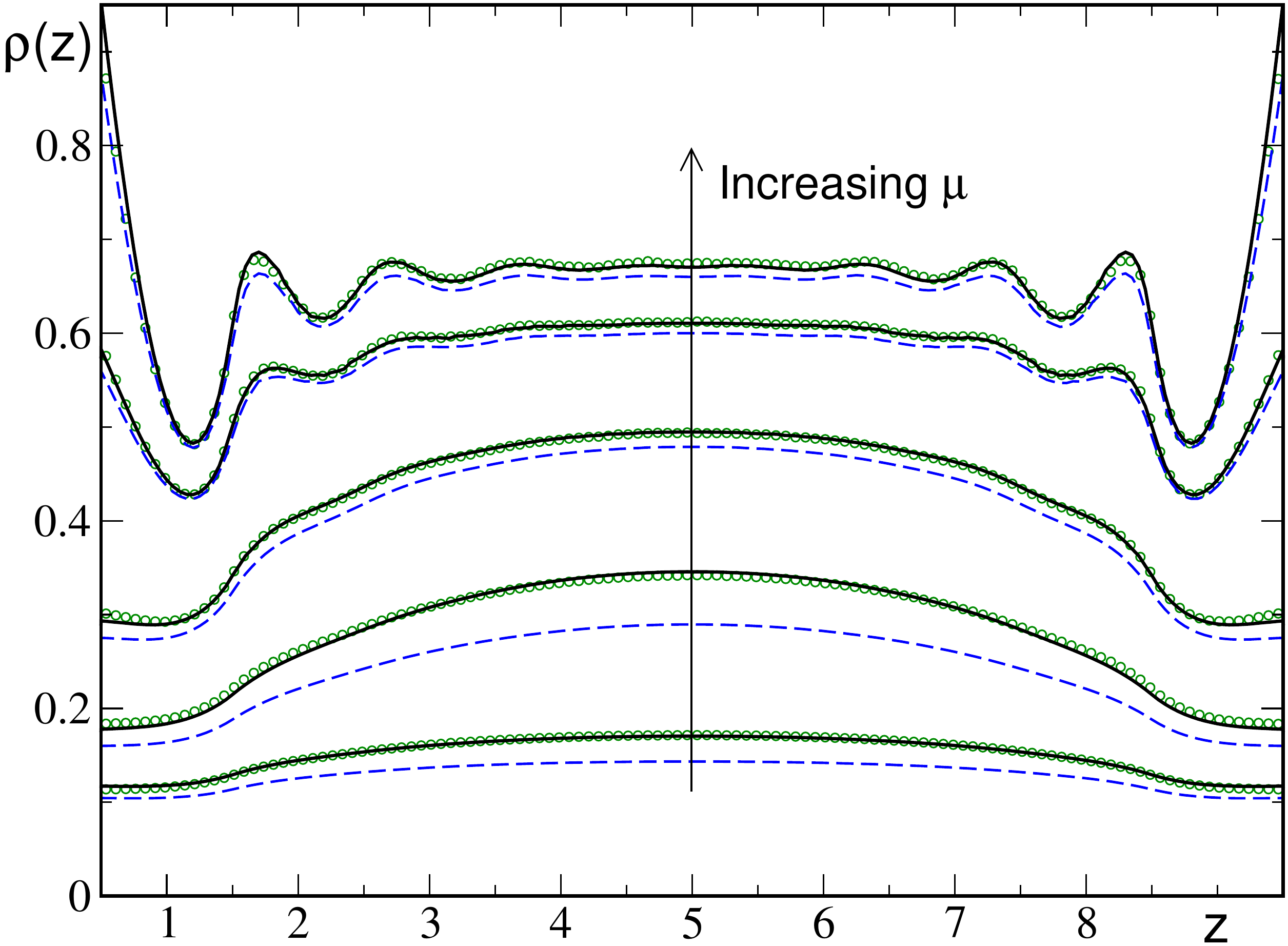}
\caption{
{\bf Planar slit.} 
The density between two hard-walls located at $z\!=\!0$ and $10$ 
for $\kappa\!=\!0.75$. 
Green circles: GCMC data at chemical potentials 
$\mu=-2.50, -2.25, -2.00, -1.50$ and $-1.00$ (moving along the arrow marked in 
Fig.\ref{fig:phase}). 
Broken blue line: BH functional at the same chemical potentials 
as used in the simulation. 
Full black line: BH functional profiles calculated by adjusting $\mu$ such that 
the average number of particles in the system matches that of simulation. 
}
\label{BHslitprofiles}
\end{figure}

 
In Fig.\ref{BHslitprofiles} we show density profiles calculated at $\kappa\!=\!0.75$ 
for five different values of the chemical potential. 
As a rule-of-thumb, if we consider the density at the centre of the gap to determine 
an effective bulk density, then tuning $\mu$ would correspond to following the 
path indicated in Fig.\ref{fig:phase}. 
For each of the five statepoints we show both the density profiles calculated at the same chemical 
potential as used in simulation (broken lines) and those calculated using the fitting 
procedure described above (full lines). We omit to show results from the SMF functional, 
because these lie so far from the simulation data that they would only serve to confuse 
the figure. For the states at $\mu\!=\!-1$ and $\mu\!=\!-1.5$ the BH functional performs very well. 
The density generated at the true chemical potential already gives a good account 
of the simulation data and only a very slight tuning of $\mu$ is required 
to create an excellent fit. 
This provides further evidence, in addition to the data shown in 
Fig.\ref{fig:radial_highrho}, that the fundamental assumption of the BH theory 
is accurate for inhomogeneous fluids at high densities. 

Deviations start to emerge as $\mu$ is reduced to lower values, reflecting the increasing 
influence of bulk critical fluctuations. 
The profile at $\mu\!=\!-2.25$ is the most affected 
by proximity to the critical point (located at $\mu^{\rm BH}_{\rm crit}\!=\!-2.47$, $\kappa^{\rm BH}_{\rm crit}\!=\!0.79$); 
the BH theory underestimates the value of the density in the centre of the gap. 
Nevertheless, for all the considered statepoints tuning $\mu$ still results in a very good fit 
to the simulation data. 
This suggests that the structural `building blocks' of the BH functional are sufficient to accurately 
describe inhomogeneous profiles at all thermodynamic statepoints and that it is rather the bulk 
thermodynamics which is insufficiently accurate in the critical region. 
It could be speculated that modifying/tuning the BH functional to have improved bulk thermodynamics, 
without increasing the structural complexity of the theory, could lead to very accurate results. 
Such an approach has been successfully applied to the original Rosenfeld hard-sphere functional to 
`upgrade' the theory from Percus-Yevick to Carnahan-Starling thermodynamics, while retaining the 
same geometrical weight functions \cite{roth}.

\subsection*{Free interface}
As a final application of the BH functional in planar geometry we consider the free interface 
between coexisting liquid and gas phases (the densities of which we will denote by $\rho_{\rm l}$ 
and $\rho_{\rm g}$). 
The nature of the density profile at the free interface 
has been the subject of much conjecture, primarily concerning the question of whether the profile 
exhibits either a monotonic or an oscillatory decay into the bulk. 
On the gas side of the profile it is established that the decay is monotonic, it is 
the decay into the bulk liquid which remains the subject of debate.

For model fluids with short-ranged interactions it can be shown that an inhomogeneous density profile 
ultimately decays into bulk in the same way as the radial distribution function of the bulk 
fluid (see Ref.~\cite{poles} and references therein). 
For the free interface this implies that if $g(r)$ exhibits damped 
oscillatory decay at the coexisting state-point on the liquid side of the binodal, then the corresponding 
liquid-vapour density profile will decay into the bulk liquid with the same frequency and decay length. 
The range of $\rho_{\rm l}$ values over which this oscillatory behavior can occur is determined by 
the point at which the binodal intersects the so-called Fisher-Widom line (a line in the 
($\rho_{\rm b},\kappa$) plane marking the cross-over from monotonic to asymptotic decay of 
$g(r\!\rightarrow\!\infty)$) \cite{fisherwidom}. 
One thus arrives at a picture in which a portion of the liquid-side of the binodal, between 
the triple point and the Fisher-Widom intersection point, should in principle be associated with 
oscillatory liquid-vapour profiles. 

\begin{figure}
\hspace*{-0.8cm}
\includegraphics[width=0.92\linewidth]{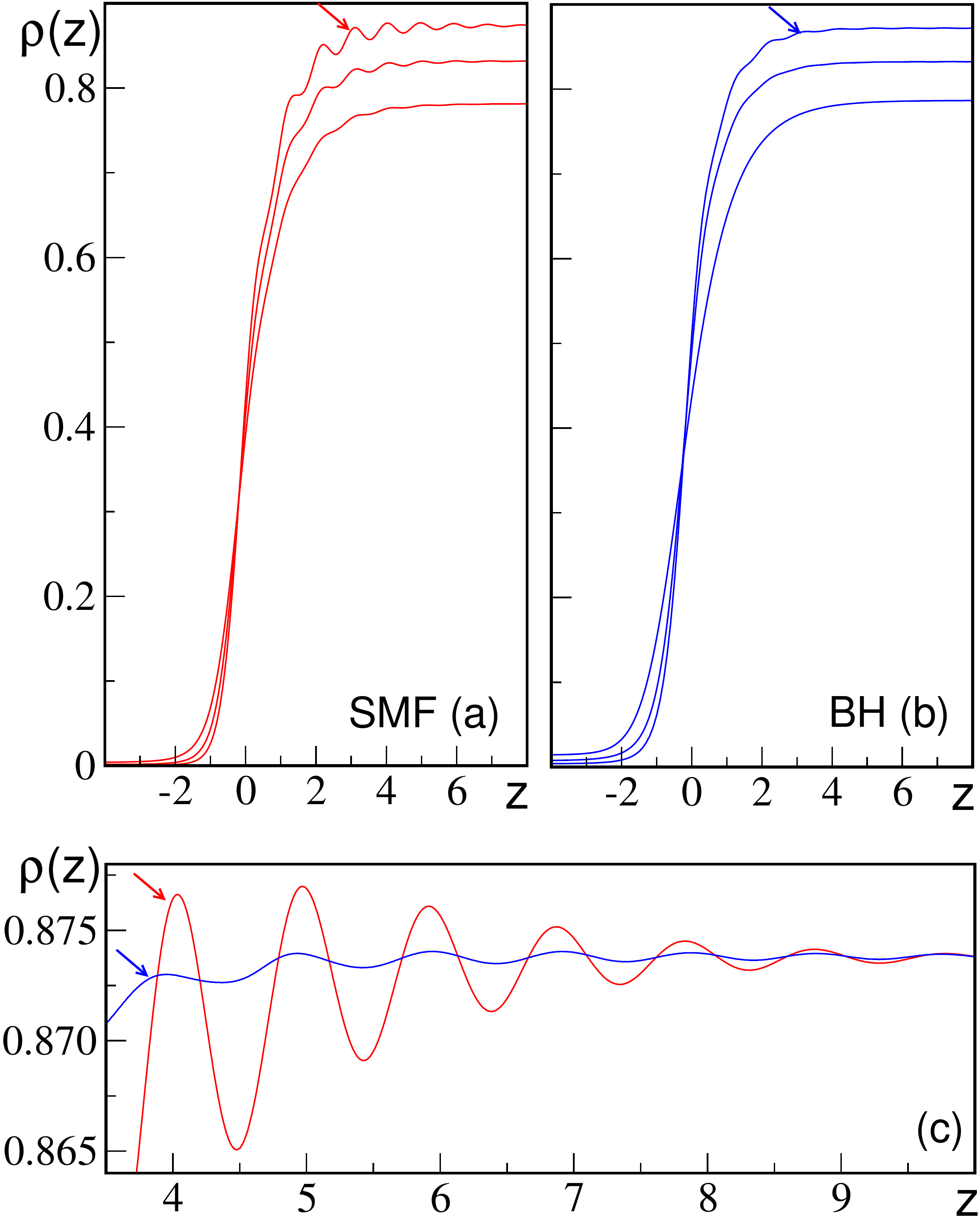}
\caption{
{\bf Free interface.} Density profiles at the liquid-vapour interface from 
(a) the SMF functional and (b) the BH functional. As the two theories 
have different binodals (see  Fig.\ref{fig:phase}) we compare profiles with equal values 
of the coexisting liquid density. Panel (c) focuses on the decay of the density 
into the bulk liquid for the statepoint closest to the triple-point. 
The arrows are intended to help the reader better see which of the oscillations 
in (a) and (b) are being shown in (c). 
\label{liquid_vapour_fig}
}
\end{figure}

An important caveat to the above arguments is that the theory of asymptotic decay makes no 
claim regarding the amplitude of any oscillatory contribution to $\rho(z)$. 
We can certainly expect any oscillations at the free interface to have smaller amplitude 
than for solid-fluid interfaces (e.g. the profiles shown in Fig.\ref{BHslitprofiles}) due 
to the influence of lateral capillary wave fluctuations. 
Previous DFT investigations have reported oscillatory profiles for 
a variety of model interaction potentials \cite{lv1,lv2,geysermans}. 
However, all of these studies employed the same SMF functional, raising the obvious question 
of whether the rather large amplitude of the observed oscillations is a generic 
feature of DFT or an artifact of the SMF approximation.  
The fact that this question has not really been addressed is a consequence of both the 
simplicity with which the SMF functional can be implemented and the lack of alternative 
approaches; 
in a sense the term 'density functional study' has become almost 
synonymous with 'standard mean-field study'. 
For this reason the BH functional is of special value, providing the first opportunity 
to test in some kind of systematic way the robustness of the SMF predictions. 

In Fig.\ref{liquid_vapour_fig} we show liquid-vapour profiles obtained from both the SMF 
and the BH functionals for values of $\kappa$ approaching the triple point. 
We refrain from presenting simulation data as this would require extensive 
computation beyond the scope of the present work (Ref.~\cite{lv_sim} gives some insight into 
the difficulties of simulating the free interface). 
As the two theories have quite different binodals 
care is required to ensure a fair comparison between the SMF and the BH density 
profiles. 
If we restrict our attention to states approaching the triple point (for which 
$\rho_{\rm l}-\rho_{\rm g}\approx\rho_{\rm l}$), 
then a reasonable comparison can be achieved by comparing profiles with equal 
values of $\rho_{\rm l}$. 
The SMF functional predicts the onset of an oscillatory profile as the triple point is 
approached, consistent with previous reports \cite{lv1,lv2,geysermans}. 
The statepoint with the highest value of $\rho_{\rm l}$ exhibits well developed oscillatory 
decay into the bulk with deviations $(\rho(z)-\rho_{\rm l})/\rho_{\rm l}$ on around the $1$\% level. 
The amplitude of these oscillations rapidly diminishes as the value of $\kappa$ 
is reduced towards the critical point.
The corresponding profiles calculated using the BH functional also show oscillatory decay 
as the triple point is approached, however, the amplitude is strongly reduced with respect 
to the SMF predictions. This is consistent with our preceeding test particle and planar-slit calculations, 
which revealed that the overestimation of attraction-induced layering structure in the SMF is 
systematically corrected by the BH theory. 
In addition, at equality of $\rho_{\rm l}$ the interfacial width predicted by the BH 
approximation is somewhat larger than that from the SMF functional. 
This observation, taken together with the reduced oscillation amplitude, suggests 
that the BH theory more accurately incorporates the influence of interfacial capillary wave fluctuations 
than the SMF theory, as we shall discuss below. 
For a careful discussion of the amplitudes and damping of oscillations at the free interface 
we direct the reader to Ref.~\cite{brader_AO}.

The development of a microscopic theory for the free interface has been a subject of renewed 
interest \cite{capillary2,resonances}. One point of progress has been the clarification of the relationship 
between the van der Waals/DFT picture, based on the density profile, and the capillary wave approach 
focused on mesoscopic fluctuations of the liquid surface \cite{mandelstam,capillary1}. 
DFT calculations for this problem generate one-dimensional density profiles that 
are completely independent of the interfacial area, $A_{\rm xy}$. 
However, computer simulation studies have shown that the free interface profile clearly depends on 
the lateral size of the simulation box, such that the profile should be considered as having 
a parametric dependence, $\rho(z)\rightarrow\rho(z;A_{\rm xy})$ \cite{lv_sim,capillary1}. 
A possible resolution of this apparent contradiction is that an approximate mean-field DFT 
only incorporates the influence of capillary-wave fluctuations up to a certain effective cutoff 
distance, $\lambda_{\rm DFT}$, within the plane of the interface.   
This provides the appealing physical picture that mean-field DFT theories are comparable 
to finite-size simulations, except that within DFT the `box size' 
is hard-wired by the specific approximation employed. 
Unfortunately, a precise statistical mechanical definition of $\lambda_{\rm DFT}$ seems to be lacking at 
present. 
 
In the absence of gravity it is well-known that mean-field theory predicts a divergent transverse 
correlation length within the interface \cite{evans92}. 
The fact that the derivative of the density profile, $d\rho(z)/dz$, remains nonvanishing indicates  
that the mean-field approximation does not sufficiently incorporate the feedback of the
long-ranged transverse correlations into the density profile. 
Our interpretation of the difference between the SMF and the BH profiles shown 
in Fig.\ref{liquid_vapour_fig} is thus that the BH theory provides a better account of the 
coupling between the one- and two-body correlation functions and thus captures this 
feedback in a more realistic fashion.






\section{Discussion}\label{discussion}

In this paper we have developed and implemented a first-principles DFT for inhomogeneous 
fluids with attractive interactions. 
The theory generalizes the successful BH bulk theory and generates  
density profiles in quantitative agreement with simulation data, marking a significant 
improvement over the commonly employed SMF functional. 
Unlike previous efforts our approach avoids reference to any bulk information and is valid for an arbitrary external 
field. 
The lack of an accurate and tractable DFT for treating attractive interactions has been a 
long-standing hinderance to theoretical progress and we hope that our findings will go some way 
towards alleviating these difficulties.
We have already mentioned the alternative theories most directly related to the BH 
functional, namely the SMF and effective density approaches 
\cite{averagerho1,averagerho2,averagerho3,averagerho4,averagerho5,averagerho6}. 
In the following we would like to discuss some additional approximation schemes which could be regarded 
as rivals to the inhomogeneous BH theory.

A central feature of the DFT formalism is the dual hierarchy of correlation functions, obtained 
by functional differentiation of the free energy \cite{evans92}. 
Differentiation with respect to an external (or interaction) potential generates 
density correlators, whereas differentiation with respect to the density generates direct 
correlation functions. 
The BH functional stems from the former way of thinking and is essentially a functional 
expansion of the excess free energy in powers of the attractive contribution to the interaction potential. 
This demands that we have a detailed understanding of a {\it given reference system} for any 
$\rho(\rv)$, which is indeed the case for hard-spheres.  
However, the problem can also be approached using the second hierarchy, by 
performing a functional Taylor expansion of the excess free energy in powers of $\rho(\rv)$ about 
a reference density \cite{evans79,evans92,rama,wu}. 
This requires that for the {\it given reference density} 
we have complete understanding of the fully interacting system, which is generally not the case. 
For this reason a bulk density is usually chosen as a reference (implicitly assuming a weakly 
nonuniform fluid) and the Taylor series is truncated at quadratic-order, such that only the bulk 
pair direct correlation functions are required as input.  
Although respectable results can be obtained for systems of repulsive particles (especially in 
the case of soft penetrable particles) the theory is less successful when applied to systems 
with an attractive component to the interaction potential. 
In particular, an excess free energy with a quadratic dependence on the density cannot describe 
two minima and is thus not capable of describing phase transitions at interfaces 
\cite{evans92}. 

The sum of all terms beyond quadratic order in the density expansion is known as the 
`bridge functional' \cite{Hansen06,attard_book}. 
Rosenfeld has shown that the quadratic functional can be much improved by replacing the true bridge 
functional of the fully interacting system with that of the hard-sphere system 
(often referred to as either the `reference functional' or `universal bridge functional' method) 
\cite{rosenfeld_bridge}. 
This approach is somewhat similar in spirit to our inhomogeneous BH theory, for which 
the total correlation function $h_{\rm hs}(\rv_1,\rv_2;[\rho])$ is assumed `universal' for 
any attractive interaction. 
On the positive side, the universal bridge functional method can make very accurate predictions 
in certain cases and has the convenient feature that only one-body functions 
are required \cite{rosenfeld_bridge,oettel,sweatman1,sweatman2}.
However, compared with the BH functional we observe two fundamental disadvantages of the 
Bridge functional approach: 
(i) Despite resumming higher order terms, the theory is still a density expansion and thus cannot escape 
the need to identify a bulk reference state. This is problematic for confined fluids. 
(ii) To yield accurate results the hard-sphere reference functional has to be evaluated at some 
effective hard-sphere diameter. This introduces a free parameter for which an optimization criterion 
must be specified. 
In our view, the assumption that $h_{\rm hs}(\rv_1,\rv_2;[\,\rho\,])$ is `universal' constitutes a 
physically clear generalization of van der Waals vision of the liquid state, whereas 
the universal bridge functional seems to be a more obscure formal object.

An alternative to the aforementioned DFT approximations is to attack the pair correlations 
directly by applying an inhomogeneous closure
to the OZ equation (e.g.~hypernetted-chain) for the {\it full} interaction potential 
\cite{attard_book}. 
%
%
While integral equation theories can provide accurate results 
(see Refs.~\cite{attard1,shoulder,kovalenko} for example) they suffer from the following well-known problems: 
(i) There exist `no-solution' regions in thermodynamic parameter-space where the 
theory fails to converge. 
In the case of liquid-gas phase separation this region typically envelopes the critical point 
and thus prevents both a proper determination of the binodal and the investigation of any 
associated interfacial phenomena. 
(ii) Thermodynamic inconsistency. 
Making an approximation on the level of the pair correlations, 
rather than on the level of the free energy, has the consequence that the density profile 
is not unique. The three formally exact routes from the pair correlations to the density will 
yield inconsistent results \cite{attard_book}. 
These failings, which are ultimately linked to the absence of a generating (free energy) 
functional, make inhomogeneous integral equations theories generally unsuitable for the 
investigation of interfacial phase transitions.  

Finally, we would like to outline some possibilities for future work. 
On the technical side, 
now that we have established the accuracy of the BH functional it would be worth to invest 
effort into improving the numerical efficiency of our algorithms. 
The bottleneck in our calculations is the iterative solution of the OZ equation, which requires 
a constant back-and-forth between real and transform (either Legendre or Hankel) space. 
We plan to investigate existing proposals to speed-up these transforms \cite{LadoSpeedup} 
as well as the possibility to generalize methods developed for solving the OZ equation 
in bulk to the inhomogeneous case (see \cite{brader_critical} and references therein). 
An alternative method to improve numerical efficiency would be to 
exploit the analytic expression for $c_{\rm hs}(\rv_1,\rv_2)$  generated by taking 
two functional derivatives of the Rosenfeld functional (c.f.~equation \eqref{derivative_def}). 

Regarding future physical investigations, the ability of the BH functional to describe 
accurately systems with an attractive component to the interaction potential could be 
exploited to address a variety of topics. Some possibilities are: 
(i) To apply the BH functional to a system interacting via a competing attractive and repulsive 
interaction (the so-called short-range attractive and long-range repulsive (SALR) class of 
potential) 
\cite{melascio,archer_wilding}. Standard liquid state theories are unable to account  
for the complex phase behavior presented by these systems and it would therefore be of interest to 
investigate the predictions of the BH functional.
(ii) There have been recent advances in obtaining {\it canonical} observables (i.e.~the density 
profile) from {\it grand-canonical} DFT; ensemble differences become important when considering 
small systems with few particles \cite{canonical transform,canonicalDDFT}. 
Application of this method to realistic systems with attractive interactions has so-far been 
hindered by the absence of an accurate grand canonical functional. The BH functional 
could thus open-up possibilities to study, e.g.~nucleation and clustering in small systems.   
(iii) Although technically challenging, it would be interesting to investigate in detail the 
inhomogeneous two-body correlations within the liquid-vapour interface, with a view to shedding 
light on the nature of the `intrinsic interface' predicted by mean-field DFT.

\section*{Appendix A}\label{app1}

Consider the functional derivative of an arbitrary functional $H$ with 
respect to a scalar function $g(\rv)$.
\begin{align}\label{test}
\frac{\delta H[g]}{\delta g(\rv)}. 
\end{align} 
The derivative can be reversed to recover $H$ 
by the following integration
\begin{align}\label{integral_example}
H[g] = H[g_{\rm r}] + \int \!d\rv \int_{g_{\rm r}(\rv)}^{g(\rv)}\!d\tilde g(\rv) 
\frac{\delta H[\tilde g]}{\delta \tilde g(\rv)}, 
\end{align}
where $g_{\rm r}(\rv)$ is a reference function and $\tilde
g(\rv)$ is a dummy integration variable. 
The integration in \eqref{integral_example} is a one-dimensional 
integral over the value of $\tilde g$ at point
$\rv$. 
Provided that the functional $H$ is unique, the result will be independent of 
the chosen path in function space. 
The simplest choice is then a linear parametric path 
\begin{align}\label{linearpath}
\tilde g(\rv)\equiv g_{\alpha}(\rv) = g_{\rm r}(\rv) + \alpha (g(\rv) - g_{\rm r}(\rv) ),
\end{align} 
where the `charging parameter' $\alpha$ varies from zero  to  unity. 
Equation \eqref{integral_example} thus becomes
\begin{align}\label{integral_example_alpha}
H[g] = H[g_{\rm r}] + \int_0^1 \!\!d\alpha \!\int \!d\rv\,
\Delta g(\rv)\frac{\delta H[g_{\alpha}]}{\delta g_{\alpha}(\rv)},   
\end{align}
where $\Delta g(\rv) \equiv g(\rv) - g_{\rm r}(\rv)$. 
Generalization to the case of two vector arguments is 
straightforward:
\begin{align}\label{example_two_arguments}
G[f] = G[f_{\rm r}] +\! \int_0^1 \!\!d\alpha \!\int \!d\rv \!\int \!d\rv'\,
\Delta f(\rv,\rv')\frac{\delta G[f_{\alpha}]}{\delta f_{\alpha}(\rv,\rv')}, 
\end{align}
where $\Delta f(\rv,\rv') \equiv f(\rv,\rv') - f_{\rm r}(\rv,\rv')$,
and we have again assumed a linear integration path.

\section*{Appendix B}\label{app2}
Our perturbative approach employs the hard-sphere reference free energy functional 
$F_{\rm hs}=F_{\rm id} + F_{\rm hs}^{\rm exc}$. 
Within Rosenfeld's original fundamental measures approach the excess Helmholtz 
free energy is given by \cite{rosenfeld89}
\begin{align}\label{ros_fe}
\beta F^{\rm exc}_{\rm hs}
[\,\rho\,] = \int d\rv_1 \; \Phi \left( \left\lbrace n_{\alpha}(\rv_1) \right\rbrace  \right),
\end{align}
where the reduced free energy density is a function of a set of weighted densities
\begin{align}
\Phi = - n_0 \ln(1-n_3) + \frac{n_1 n_2 - {\bf n}_1 \cdot {\bf n}_2}{1-n_3} + \frac{n_2^3 
- 3 n_2 {\bf n}_2 \cdot {\bf n}_2}{24 \pi (1-n_3)^2}.
\end{align}
The four scalar weighted densities, $n_0\cdots n_3$, and two vector weighted densities, 
${\bf n}_1$ and ${\bf n}_2$, are given by 	
\begin{align}
n_{\alpha}(\rv_1) = \int d\rv_2 \; \rho(\rv_2)\, \omega_{\alpha}(\rv_1-\rv_2), 
\end{align}
where the weight functions, characteristic of the geometry of the hard-spheres 
with a radius $R$, are given by 
\begin{align}
\omega_3(\rv)&=\Theta(R-r), \notag\\
\omega_2(\rv)&=\delta(R-r), \notag\\
\boldsymbol{\omega}_2(\rv)&=\frac{\rv}{r}\delta(R-r),
\end{align}
and $\omega_1(\rv)=\omega_2(\rv)/(4\pi R)$, 
$\omega_0(\rv)=\omega_2(\rv)/(4\pi R^2)$ 
and $\boldsymbol{\omega}_1(\rv)=\boldsymbol{\omega}_2(\rv)/(4\pi R)$. 
The (negative) first functional derivative of the excess free energy yields 
the one-body direct correlation function
\begin{align}
c^{(1)}_{\rm hs}(\rv_1)=-\sum_{\alpha}\int d\rv_2 
\frac{\partial \Phi}{\partial n_{\alpha}(\rv_2)}\omega_{\alpha}(\rv_2-\rv_1).
\end{align}
Explicit expressions for $c^{(1)}_{\rm hs}(\rv_1)$ in both planar and spherical geometry 
can be found in subsections 8.2 and 8.3 of Ref.~\cite{roth}.

\section*{Appendix C} 
For the hard-sphere system both $h_{\rm hs}(\rv_1,\rv_2)$ and $c_{\rm hs}(\rv_1,\rv_2)$ are 
discontinuous when $|\rv_1-\rv_2|=d$. If these functions are directly transformed then the precise 
location of the discontinuity becomes uncertain on the order of the numerical grid spacing. 
Using diagrammatic analysis it can be shown that the function 
\begin{align}\label{gamma_def}
\gamma(\rv_1,\rv_2) = h(\rv_1,\rv_2)  - c(\rv_1,\rv_2)  
\end{align}
is a continuous function for any interaction potential \cite{attard_book}. 
Using equation \eqref{gamma_def} to eliminate $h_{\rm hs}(\rv_1,\rv_2)$ from the OZ equation 
\eqref{oz} yields an alternative form
\begin{align}\label{alternative_form}
\gamma_{\rm hs}(\rv_1,\rv_2) &= 
\int\! d\rv_3 \,c^{}_{\rm hs}(\rv_1,\rv_3)\rho(\rv_3) c_{\rm hs}^{}(\rv_3,\rv_2)
\notag\\
&+ 
\int\! d\rv_3\, \gamma^{}_{\rm hs}(\rv_1,\rv_3)\rho(\rv_3) c_{\rm hs}^{}(\rv_3,\rv_2),
\end{align}
for which we only have to deal with one discontinuous function, namely $c_{\rm hs}$. 
In spherical geometry a Legendre transformation reduces equation \eqref{alternative_form} 
to an equation for the transforms
\begin{align}\label{alternative_form_sph}
G_n(r_1,r_2) &= \frac{4\pi}{2n+1}\Bigg(\int_0^{\infty}\!\!dr_3\,r_3^2\,C_n(r_1,r_3)\rho(r_3)\,C_n(r_3,r_2). 
\notag
\\ 
&+\!\! \int_0^{\infty}\!\!dr_3\,r_3^2\,G_n(r_1,r_3)\rho(r_3)\,C_n(r_3,r_2) \Bigg), 
\end{align}
where $G_n$ is the Legendre transform of $\gamma_{\rm hs}$. 
In planar geometry a Hankel transform of equation \eqref{alternative_form} generates the following 
simplified form 
\begin{align}\label{alternative_form_planar}
\mathcal{G}_{k}(z_1,z_2) &= \int_{-\infty}^{\infty} \!\!dz_3 \, 
\mathcal{C}_{k}(z_1,z_3)\rho(z_3)\mathcal{C}_{k}(z_3,z_2)
\notag\\
&+
\int_{-\infty}^{\infty} \!\!dz_3 \, 
\mathcal{G}_{k}(z_1,z_3)\rho(z_3)\mathcal{C}_{k}(z_3,z_2),
\end{align}
where $\mathcal{G}_{k}$ is the Hankel transform of $\gamma_{\rm hs}$. 
The general equation \eqref{oz_epsilon} for the perturbed pair correlations as well as  
the reduced forms for spherical and planar geometry, equations \eqref{oz_sph_pert} 
and \eqref{oz_planar_epsilon}, respectively, can also be rewritten in the alternative form 
by trivial extension of the above expressions.

Equations \eqref{alternative_form_sph} and \eqref{alternative_form_planar} both require 
a discrete integral transform of the discontinuous pair direct correlation function. 
However, the analogous procedure for the Hankel transform has not been documented. 
Defining a critical radius  $R_{\rm c}=(d^2 - (z_1 -z_2)^2)^{\frac{1}{2}}$ we can at any time in 
the iterative cycle use the continuous function $\gamma_{\rm hs}$ to evaluate 
both the direct correlation function 
\begin{align}
c_{\rm hs}(z_1,z_2,R_{\rm c}) &= - 1 - \gamma_{\rm hs}(z_1,z_2,R_{\rm c})
\notag\\
&= c_{\rm st}(z_1,z_2),
\end{align}
and the derivative
\begin{align}
\frac{\partial c_{\rm hs}(z_1,z_2,\bar{r}_{12})}{\partial \bar{r}_{12}}
{\bigg|}_{R_{\rm c}} 
&= 
-\frac{\partial \gamma_{\rm hs}(z_1,z_2,\bar{r}_{12})}{\partial \bar{r}_{12}}
{\bigg|}_{R_{\rm c}}
\notag\\
&= c_{\rm sl}(z_1,z_2). 
\end{align}
We then use these quantities to define the following linear step function
\begin{align}\label{stepslope}
f(z_1,z_2,\bar{r}) = \begin{cases}
c_{\rm st} 
+ c_{\rm sl}(\bar{r}-R_c),  &\bar{r}<R_c \\
\hspace*{0.55cm}0 ,                    &\bar{r}>R_c
\end{cases}
\end{align}
which will allow us to remove the unwanted discontinuity.  
The analytical Hankel transform of equation \eqref{stepslope} is given by
\begin{align}
\overline{f}(k) &= c_{\rm st} \frac{2\pi R_c}{k} J_1(kR_c) 
\\
&- c_{\rm sl} 
\Bigg(
\frac{\pi^2R_c}{k^2}\Big(J_1(kR_c)S_0(kR_c)-J_0(kR_c)S_1(kR_c)\Big)
\Bigg),
\notag
\end{align}
where $S_0$ and $S_1$ are the zeroth and first-order Struve functions, respectively. 
Thus, to numerically Hankel transform $c_{\rm hs}$ we perform the following steps: 
(i) Construct the continuous and smooth function $\alpha = c_{\rm hs}\!-\!f$, 
(ii) Numerically transform to obtain $\bar{\alpha}$, 
(iii) Add the analytic transform, $\bar{c}_{\rm hs}=\bar{\alpha}+\bar{f}$. 
To perform the inverse transform we simply reverse this procedure: 
(i) Construct $\bar{\alpha}=\bar{c}_{\rm hs}-\bar{f}$, 
(ii) Numerically inverse transform to get $\alpha$, 
(iii) Add the linear step, $c_{\rm hs} = \alpha + f$. 
Using these techniques the numerical transform is at no point confronted with a 
discontinuous function.
An analogous treatment of the Legendre transform for the case of spherical symmetry 
is described in the Appendix of Ref.~\cite{attard1}.


\begin{thebibliography}{}

\bibitem{vdw}
J.D.~van der Waals, thesis, Univ. Leiden (1873).

\bibitem{boltzmann}
L.~Boltzmann, {\it Lectures on Gas Theory} (translated by S.G.~Brush),
(Berkeley, 1964).

\bibitem{ornstein}
L.S.~Ornstein and F.~Zernike, Proc. Acad. Sci. Amst., 
{\bf 17}, 793 (1914).

\bibitem{kac} 
M.~Kac, G.E.~Uhlenbeck and P.C.~Hemmer, J. math. Phys.,
{\bf 4}, 216, 229 (1963);
M.~Kac, G.E.~Uhlenbeck and P.C.~Hemmer, J. math. Phys., 
{\bf 5}, 60 (1964).

\bibitem{zwanzig}
R.W.~Zwanzig, J. Chem. Phys. {\bf 22}, 1420, (1954).

\bibitem{py}
J.K.~Percus and G.J.~Yevick, Phys. Rev. {\bf 110}, 1 (1958).

\bibitem{wertheim}
M.S.~Wertheim, Phys. Rev. Lett. {\bf 10} 321 (1963).

\bibitem{thiele}
E.~Thiele, J. Chem. Phys. {\bf  39}, 474 (1963).

\bibitem{smith}
W.~R. Smith and D. Henderson,
Mol. Phys. {\bf 19}, 411 (1970).

\bibitem{bh_original1}
J.~A. Barker and D. Henderson,
J. Chem. Phys., {\bf 47}, 2856 (1967).

\bibitem{bh_original2}
J.~A. Barker and D. Henderson,
J. Chem. Phys., {\bf 47}, 4714 (1967).

\bibitem{bh_review}
J.~A. Barker and D. Henderson,
Rev. Mod. Phys. {\bf 48}, 587 (1976).


\bibitem{scoza}
J.~S. H\o ye and G. Stell, J. Chem. Phys., {\bf 67} 439 (1977).

\bibitem{stell} 
D. Pini, G. Stell and N.~B. Wilding, Mol. Phys., {\bf 95}, 483 (1998).

\bibitem{hrt}
A. Parola and L. Reatto, Phys. Rev. Lett. {\bf 53} 2417 (1984); 
A. Parola and L. Reatto, Phys. Rev. A {\bf 31} 3309 (1985).

\bibitem{Hansen06}
J.~P.~Hansen and I.~R.~McDonald, {\it Theory of Simple Liquids}, 3rd ed. 
(Academic Press, London, 2006).

\bibitem{evans79}
R.~Evans, Adv. Phys. {\bf 28},  143  (1979).

\bibitem{evans92}
R.~Evans, {\it Density functionals on the theory of nonuniform fluids}. In {\it 
Fundamentals of Inhomogeneous Fluids}, Edited by D. Henderson (Dekker, New York 1992).

\bibitem{averagerho1}
M. H. Kalos, J. K. Percus and M. Rao, 
J. Stat. Phys., {\bf 17}, (1977). 

\bibitem{averagerho2}
S. Sokolowski and J. Fischer, 
J. Chem. Phys. {\bf 96}, 5441 (1992).

\bibitem{averagerho3}
J. C. Barrett, 
J. Chem. Phys. {\bf 124}, 144705 (2006).

\bibitem{averagerho4}
P. Lurie-Gregg, J. B. Schulte and D. Roundy, 
Phys. Rev. E {\bf 90} 042130 (2014).

\bibitem{averagerho5}
Z. Tang, L.E. Scriven and H.T. Davis, 
J. Chem. Phys. {\bf 95} 2659 (1991).

\bibitem{averagerho6}
S. Varga, D. Boda, D. Henderson and S. Sokolowski, 
J. Colloid and Interface Science, {\bf 227} 223 (2000). 

\bibitem{chacko}
A.~J. Archer, B. Chacko and R. Evans,
J. Chem. Phys. {\bf 147}, 034501 (2017)

\bibitem{line_integral}
J.~M. Brader and M. Schmidt
Mol. Phys. {\bf 113} 2873 (2015).

\bibitem{sullivan}
D.E. Sullivan, 
Phys. Rev. A {\bf 25} 1669 (1982).

\bibitem{orpa}
H.C.~Anderson, D.~Chandler and J.D.~Weeks, 
J.Chem.Phys. {\bf 56} 3812 (1972).

\bibitem{smith2}
B.~D. Kelly, W.~R. Smith and D. Henderson,
Mol. Phys. {\bf 114}, 2446 (2016).


\bibitem{rosenfeld89}
Y.~Rosenfeld, 
Phys. Rev. Lett. {\bf 63} 980 (1989).

\bibitem{roth}
R.~Roth, J.Phys.:Condens.Matter {\bf 22} 063102 (2010).  

\bibitem{attard_book}
P. Attard, {\it Thermodynamics and statistical mechanics: 
Equilibrium by Entropy Maximisation} (Elsevier, (2002)).

\bibitem{attard1}
P. Attard, 
J. Chem. Phys. {\bf 91}, 3072 (1989).

\bibitem{Allen}
M.P. Allen and D. J. Tildesley, {\it Computer simulation of liquids} 
(Oxford university press, (2017)).

\bibitem{lado}
F. Lado, 
J. Comp. Phys. {\bf 8} 417 (1971). 

\bibitem{triple}
Simulations of the full phase diagram for values of $\alpha\ge 3.9$ have been performed 
by L. Mederos and G. Navascu\'ez,
J. Chem. Phys. {\bf 101}, 9841 (1994). Extrapolation of their data allows us to roughly 
estimate the triple point density for $\alpha=1.8$.


\bibitem{poles}
R. Evans, R. J. F. Leote de Carvalho, J. R. Henderson and D. C. Hoyle, 
J. Chem. Phys. 100, {\bf 591} (1994).

\bibitem{fisherwidom}
M.E. Fisher and B. Widom, 
J. Chem. Phys. {\bf 50}, 3756 (1969).



\bibitem{lv1}
R. Evans, J. R. Henderson, D. C. Hoyle, A. O. Parry and Z. A. Sabeur, 
Mol. Phys. {\bf 80} 755 (1993).

\bibitem{lv2}
R. Checa, E. Chac\'on and P. Tarazona, 
Phys. Rev. E {\bf 70} 061601 (2004).

\bibitem{geysermans}
P.~Geysermans, N.~Elyeznasni and V.~Russier,
J. Chem. Phys. {\bf 123}, 204711 (2005). 




\bibitem{lv_sim}
P. Tarazona, E. Chac\'on, M. Reinaldo-Falag\'an and E. Velasco,  
J. Chem. Phys. {\bf 117} 3941 (2002).

\bibitem{brader_AO}
J.M. Brader, R.Evans and M.Schmidt, Mol.Phys. 101 (23-24), 3349 (2003)

\bibitem{capillary2}
E.M. Fern\'andez E. Chac\'on, P. Tarazona, A.O. Parry and C. Rasc\'on, 
Phys. Rev. Lett. {\bf 111} 096104 (2013).

\bibitem{resonances}
A.O. Parry and C. Rasc\'on, 
Nature Physics {\bf 15} 287 (2019).

\bibitem{mandelstam}
J.S. Rownlinson and B. Widom, {\it Molecular Theory of Capillarity} (Oxford: Clarendon). 

\bibitem{capillary1}
P. Tarazona, E. Chac\'on and F. Bresme, 
J. Phys.: Condens. Matter {\bf 24} 284123 (2012).




\bibitem{rama}
T. Ramakrishnan and M. Yussouff, 
Phys. Rev. B {\bf 19} 2775 (1979).

\bibitem{wu}
Y. Tang and J. Wu, 
Phys. Rev. E {\bf 70} 011201 (2004).

\bibitem{rosenfeld_bridge}
Y.~Rosenfeld, J. Chem. Phys. {\bf 98}, 8126 (1993).

\bibitem{oettel}
M. Oettel, 
J. Phys.: Condens. Matter {\bf 17} 429 (2005).

\bibitem{sweatman1}
M.B. Sweatman, 
PhD thesis, University of Bristol, UK, (1995).

\bibitem{sweatman2}
M.B. Sweatman, 
Mol. Phys. {\bf 98}, 573 (2000).

\bibitem{shoulder}
J.M. Brader, 
J. Chem. Phys. {\bf 128}, 104503 (2008).

\bibitem{kovalenko}
I. Omelyan, F. Hirata and A. Kovalenko, 
Phys. Chem. Chem. Phys., {\bf 7}, 4132 (2005).

\bibitem{LadoSpeedup}
F. Lado, 
Mol. Phys. {\bf 107} 301 (2009). 

\bibitem{brader_critical}
J.M. Brader, 
International Journal of Thermophysics {\bf 27}, 394 (2006). 

\bibitem{melascio}
G. Melascio, 
J. Phys. Condens. Matter {\bf 19} 073101 (2007).

\bibitem{archer_wilding} 
A.J. Archer and N.B. Wilding, 
Phys. Rev. E, {\bf 76}, 031501 (2007).

\bibitem{DDFT}
U.M.B. Marconi and P. Tarazona, 
J. Chem. Phys. {\bf 110}, 8032 (1999).

\bibitem{canonical transform}
D. de las Heras, and M. Schmidt
Phys. Rev. Lett., {\bf 113}, 238304, (2014). 

\bibitem{canonicalDDFT}
T. Schindler, R. Wittmann and J.M. Brader, 
Phys.Rev.E {\bf 99} 012605 (2019). 

\end{thebibliography}
\end{document}